# Exploring Multi-Reader Buffers and Channel Placement during Dataflow Network Mapping to Heterogeneous Many-core Systems

**MARTÍN LETRAS, JOACHIM FALK, and JÜRGEN TEICH (Fellow, IEEE)**
Friedrich-Alexander-Universität Erlangen-Nürnberg (FAU), Erlangen, Germany (e-mail:{martin.letras, joachim.falk, juergen.teich}@fau.de)
Corresponding author: Martin Letras (e-mail: martin.letras@fau.de).

**ABSTRACT** This paper presents an approach for reducing the memory requirements of dataflow applications, particularly from the image and video processing domain, while maximizing the throughput (minimizing the execution period) when deployed on a many-core target. Often, straightforward implementations of dataflow applications suffer from data duplication if identical data has to be processed by multiple actors. In fact, multi-cast (also called fork) actors can produce huge memory overheads when storing and communicating copies of the same data. As a remedy, so-called Multi-Reader Buffers (MRBs) can be utilized to forward identical data to multiple actors in a FIFO manner while storing each data item only once by sharing. However, using MRBs may increase the achievable period due to communication contention when accessing the shared data. This paper proposes a novel multi-objective design space exploration approach that selectively replaces multi-cast actors with MRBs and explores actor and FIFO channel mappings to find trade-offs between the objectives of period, memory footprint, and core cost. In distinction to the state-of-the-art, our exploration approach considers (i) memory-size constraints for on-chip memories, (ii) hierarchical memories to implement the buffers, e.g., tile-local memories, (iii) supports heterogeneous many-core platforms, i.e., core-type dependent actor execution times, and (iv) optimizes the buffer placement and overall scheduling to minimize the execution period by proposing a novel combined actor and communications scheduling heuristic for period minimization called CAPS-HMS. Our results show that the explored Pareto fronts improve a hypervolume indicator over a reference approach by up to 66 % for small to mid-size applications and 90 % for large applications. Moreover, selectively replacing multi-cast actors with corresponding MRBs proves to be always superior to never or always replacing them. Finally, it is shown that the quality of the explored Pareto fronts does not degrade when replacing the efficient scheduling heuristic CAPS-HMS by an exact integer linear programming (ILP) solver that requires orders of magnitude higher solver times and thus cannot be applied to large dataflow network problems.

## I. INTRODUCTION

Modern many-core systems provide ample computational power due to a large number of available cores. To exploit the available number of cores, applications should exhibit sufficient concurrency to fully utilize all cores. Imperative programming languages are often considered poorly suited for developing concurrent applications [1]. Hence, applications should be specified using a Model of Computation (MoC) that explicitly expresses concurrency, e.g., using a dataflow MoC [2], where an application is represented by a Dataflow Graph (DFG). DFG vertices represent *actors*, and edges represent *First In First Out (FIFO) channels* transmitting *tokens*. Actors thereby specify an application's computations. The dynamics of a dataflow graph is given by the notion of firings. An actor is called enabled for firing (execution) if enough tokens have accumulated at its input channels. Per firing, it consumes tokens on its input channels and produces tokens at its outputs according to a set of *firing rules*. In so-called *marked graphs* [3], the firing rules are that at least one token must exist at each input of an actor to be enabled for firing. Per firing, it consumes exactly one token on each of its inputs and produces exactly one token at each of its outputs.

One application domain well suited to use dataflow modeling is image processing. Generally, an image processing application consists of a graph of image processing filters, where each filter operates on its input and produces transformed image data at its outputs. Each filter of an image processing application can be naturally modeled by an actor.

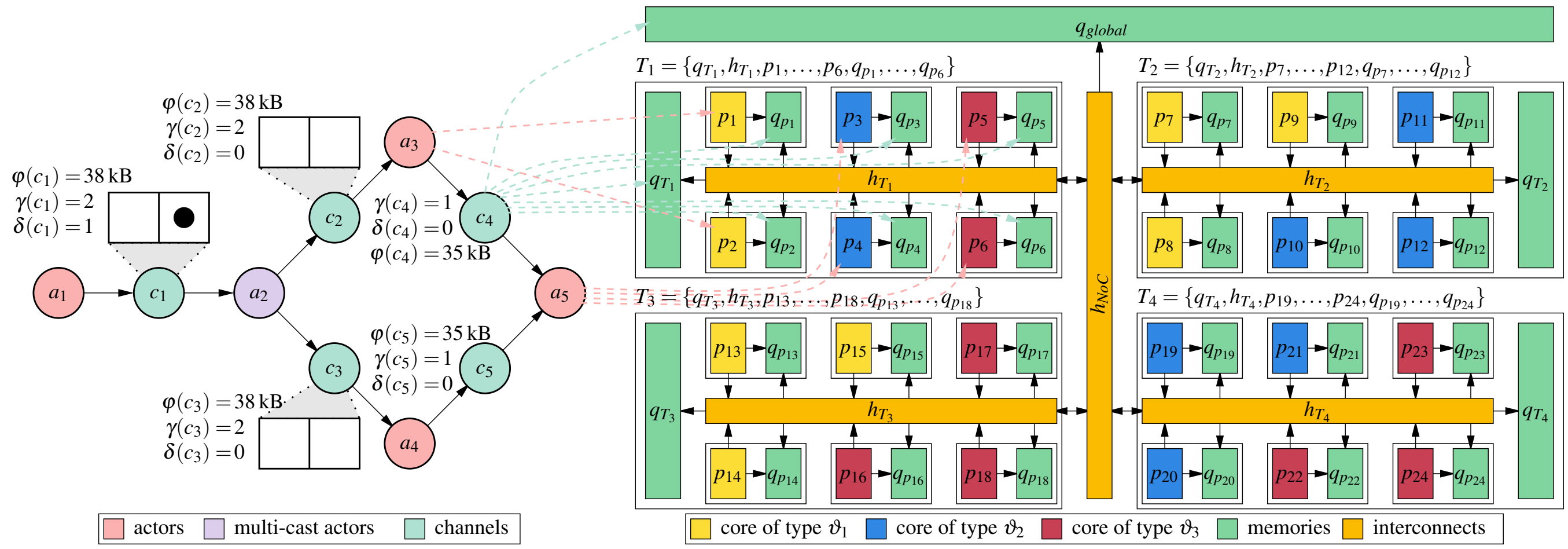

**FIGURE 1.** On the left, an application graph $g_A$ consisting of a set of actors $a_i \in A$ communicating over a set of FIFO channels $c_j \in C$ is shown. Channel capacities in terms of tokens $\gamma(c_j)$ are illustrated by white boxes. The token size in bytes $\varphi(c_j)$ and the number of initial tokens $\delta(c_j)$, e.g., one initial token for channel $c_1$ (black dot), is also illustrated. On the right, a heterogeneous four-tile many-core architecture is modeled by an architecture graph $g_R$. Processor cores are denoted $p_i$, and tiles are denoted $T_j$. Each core $p_i \in P$ can, in principle, access any core-local memory $q_{p_j} \in Q_P$, any tile-local memory $q_{T_j} \in Q_T$, as well as the global memory $q_{global}$. Dashed arcs represent mapping options from actors to cores and channels to memories. To exemplify, mapping edges are illustrated for the actors $a_3$ and $a_5$ as well as the channel $c_4$. However, to reduce visual clutter, only the resources of tile $T_1$ and the global memory are shown as targets for these mappings. In the proposed approach, actors can be mapped in principle (light red arcs) to all cores of a type that supports the execution of the actor, e.g., cores of type $\vartheta_1$ for actor $a_3$ and cores of type $\vartheta_2$ or $\vartheta_3$ for actor $a_4$. In contrast, channels can generally be mapped (light green arcs) to any memory.

In order to map a dataflow graph such as exemplified in Fig. 1 (left) with explicit modeling of actors and channels onto a many-core target such as shown in Fig. 1 (right), the actors must be properly mapped to individual cores, and the channels must be mapped to proper memories of the target architecture [4, 5]. Moreover, a schedule needs to be determined for the actor executions as well as the transport of data from and to the allocated channel memories so as to achieve a short period on the one hand while reducing the required memory footprint and core count on the other hand.

One problem of dataflow MoCs, however, is that it does not allow multiple actors to read data from the same channel. Instead, multiple individual channels must be created for the multiple readers, thus creating copying and data overheads by introducing so-called *multi-cast actors*. The only purpose of these multi-cast actors is to read the data from the producer and copy it to all consumer actors [6–8]. An example of such a multi-cast actor is the actor $a_2$ highlighted in Fig. 1 (left). Apart from high memory footprint requirements, these actors also typically cause a huge amount of communication.

To avoid copying and the resulting data duplication, [9] recently introduced a concept called *Multi-Reader Buffer (MRB)*, which is a channel that has one writer and multiple readers that behave as if each reader has a dedicated channel of the very same channel data but stores each token in the channel only once, e.g., see Fig. 2b. A minimal memory footprint for an application results when each multi-cast actor (and their connected FIFOs) is replaced by an MRB. But, as was stated in [9], this buffer replacement scheme may impact the minimum achievable period. Unfortunately, the approach presented in [9] is also limited in (i) being restricted to simple bus-based homogeneous architectures and in (ii) ignoring the capacities of the on-chip memories. Often, many-core systems, particularly Multi-Processor Systems-on-a-Chip (MPSoCs), are comprised of different core types and have dozens of cores with constrained on-chip memories connected via a hierarchical interconnect, e.g., see Fig. 1 (right).

Coping with these deficiencies, this paper contributes a multi-objective Design Space Exploration (DSE) approach that (i) considers memory-size constraints for all on-chip memories, (ii) explicitly models memory hierarchies, (iii) supports heterogeneous many-core platforms, i.e., core-type dependent actor execution times, and (iv) optimizes the buffer placement and overall scheduling to minimize the period (i.e., maximizing application throughput) by proposing a novel modulo scheduling-based heuristic named Communication-Aware Periodic Scheduling on Heterogeneous Many-core Systems (CAPS-HMS). CAPS-HMS periodically schedules actors on cores and read/write operations on hierarchical interconnect topologies in a very efficient way. To analyze the trade-off between memory footprint and period, our exploration also *selectively* replaces multi-cast actors by MRBs and explores actor and FIFO channel mappings on top of finding a periodic actor and memory access schedule. In addition to the minimization of the memory footprint and the period, the (weighted) number of allocated cores is also minimized.

The paper is structured as follows: Section II presents the fundamental models of applications and architectures and

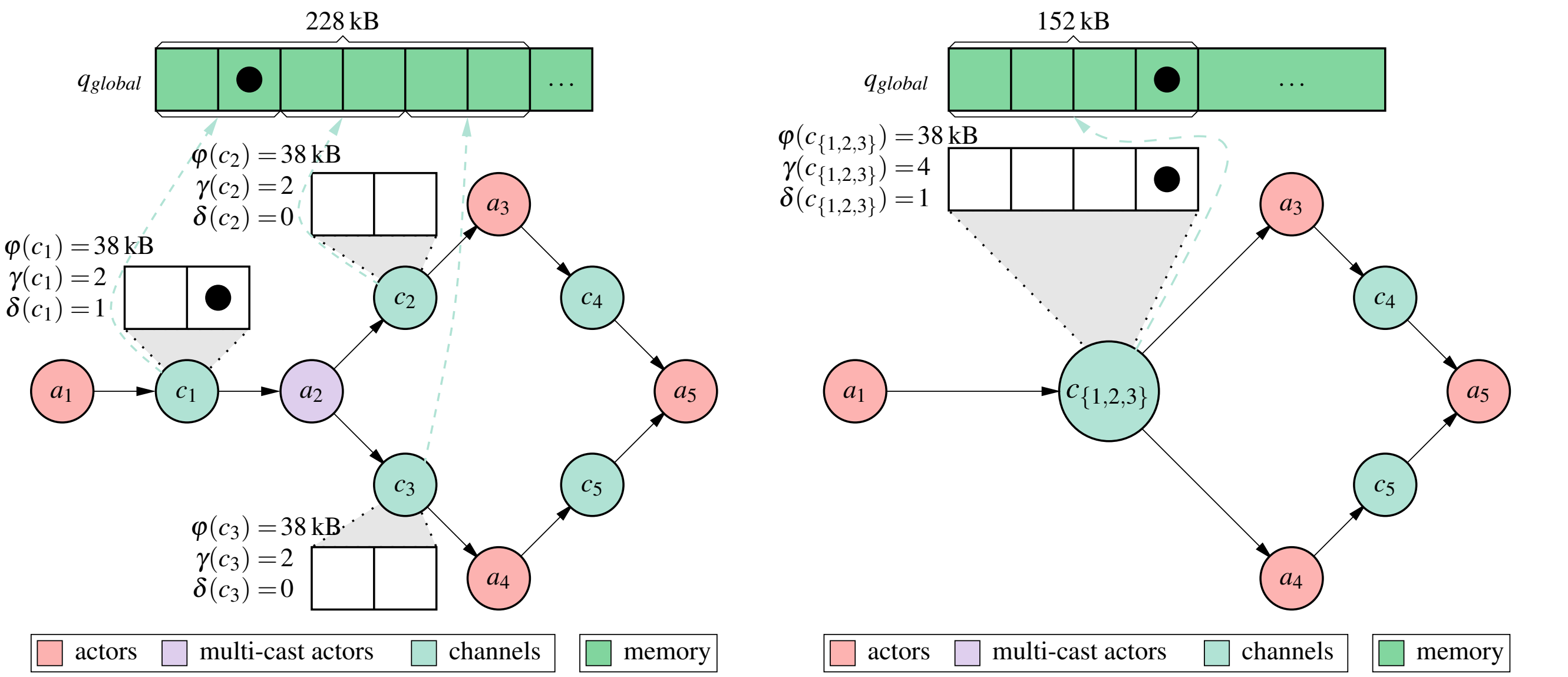

**(a)** Channels $c_1$, $c_2$, and $c_3$ allocated in the global memory. **(b)** MRB $c_{\{1,2,3\}}$ replacing multi-cast actor $a_2$ and its adjacent channels.

**FIGURE 2.** Concept of Multi-Reader Buffers (MRBs): For an application graph shown in (a), each channel connected to the multi-cast actor $a_2$ is realized as a FIFO allocated in this example in the global memory $q_{global}$. In (b), memory minimization is performed by merging the channels $c_1, c_2$, and $c_3$ into a single MRB $c_{\{1,2,3\}}$. As can be seen, the memory requirements for all channels $c_1, c_2$, and $c_3$ are identical, and each evaluates to $\gamma(c) \cdot \varphi(c) = 2 \cdot 38\,\text{kB} = 76\,\text{kB}$, resulting in an overall memory footprint of $3 \cdot 76\,\text{kB} = 228\,\text{kB}$. In contrast, the MRB $c_{\{1,2,3\}}$ replacing the channels $c_1, c_2$, and $c_3$ and the multi-cast actor $a_2$ only requires $\gamma(c_{\{1,2,3\}}) \cdot \varphi(c_{\{1,2,3\}}) = 4 \cdot 38\,\text{kB} = 152\,\text{kB}$. The channel capacity of four tokens for the MRB $c_{\{1,2,3\}}$ can be derived from the observation that across the two FIFOs (i.e., $c_1$ and $c_2$) connecting actor $a_1$ to $a_3$ at most four tokens can ever accumulate. The same holds for the FIFOs connecting actor $a_1$ to $a_4$, i.e., $\gamma(c_{\{1,2,3\}}) = \gamma(c_1) + \gamma(c_2) = \gamma(c_1) + \gamma(c_3) = 4$.

introduces the notion of multi-cast actors, all forming a so-called *specification graph* that serves as input to our optimization problem. Then, Section III introduces the design space of selective MRB replacements, actor and channel bindings, as well as actor and communication scheduling. In Section IV, a population-based DSE approach is presented in which a Multi-Objective Evolutionary Algorithm (MOEA) is used to explore the design space to find Pareto-optimal implementations. Section V presents two alternative approaches for the combined periodic actor and buffer access scheduling problem: (i) an exact formulation based on an integer linear program and (ii) our novel scheduling heuristic called CAPS-HMS for the combined periodic scheduling of actors and communications between actors. Even though the Integer Linear Program (ILP) modulo scheduling approach performs well in terms of solution times for small to mid-sized applications, the CAPS-HMS heuristic provides superior results when tackling large applications because the ILP runs into timeouts, or solution times would become prohibitively long. To evaluate the overall approach, experimental results are reported for three applications in terms of the quality of the found non-dominated sets of solutions in Section VI. It is shown for both CAPS-HMS and the ILP alternative, improvements in the quality of found solutions can be achieved when selectively replacing the multi-cast actors with MRBs when exploring the mappings of a dataflow streaming application to heterogeneous many-core architectures. In particular, for small to mid-size applications, the reported improvements range from 28 % to 66 % in the hypervolume score. In contrast, for the largest benchmark application, the reported improvement is even 90 % of the same hypervolume score. Moreover, when comparing the Pareto front quality of the CAPS-HMS heuristic against the front achieved using the exact ILP approach, the observed degradation of CAPS-HMS turns out to be minor for all presented test applications. It will be shown that for the small and mid-sized applications used in the experiments, CAPS-HMS is slightly inferior by just 7 % in terms of hypervolume compared to the ILP. Particularly for large applications and with increasing complexity of the target architecture, the ILP solution times turn out to become prohibitively long. In contrast, the fast CAPS-HMS outperforms the ILP by 67 % in hypervolume for our large test application. Section VII presents related work, and Section VIII concludes the paper.

## II. FUNDAMENTALS

The problem of mapping applications to many-core targets is often described by a *specification graph* [5, 10, 11] composed of (i) an *application graph*, (ii) an *architecture graph*, and (iii) a set of *mapping edges* that will be explained in the following.

### A. APPLICATION GRAPH

An application is modeled as a bipartite graph of actors and channels, called an *application graph*, as defined below:

*Definition 2.1 (Application Graph):* An application graph $g_A = (A \cup C, E)$ is a bipartite graph with its vertices parti-tioned into a *set of actors* $A$ and a *set of channels* $C$. Such an application graph can be derived from a DFG by explicitly modeling the FIFO channels as vertices. The *delay* function $\delta : C \rightarrow \mathbb{N}_0$, *capacity* function $\gamma : C \rightarrow \mathbb{N}$, and *size* function $\varphi : C \rightarrow \mathbb{N}$, respectively, assign each channel a number of initial tokens, a maximal number of tokens that can be stored, and the token size in bytes. The set of directed edges $E = E_\mathrm{O} \cup E_\mathrm{I}$ describes the flow of data between actors and channels and is partitioned into actor outgoing ($E_\mathrm{O} \subseteq A \times C$) and actor incoming ($E_\mathrm{I} \subseteq C \times A$) edges. Throughout this paper, we assume marked graph semantics [3] of the application graph. Finally, the function $\tau : A \times \Theta \rightarrow \mathbb{N} \cup \{\perp\}$ represents the execution time $\tau(a, \vartheta)$ of an actor $a$ when mapped on a core of type $\vartheta \in \Theta$. The $\perp$ value indicates that an actor $a$ cannot be mapped to a particular core type $\theta$.

In Figs. 1 and 2a, an example of an application graph $g_\mathrm{A}$ consisting of five actors $A = \{a_1, \ldots, a_5\}$ communicating via five channels $C = \{c_1, \ldots, c_5\}$ is given. Each communication channel $c \in C$ has annotated its corresponding number of initial tokens $\delta(c)$, capacity $\gamma(c)$, and size of each token $\varphi(c)$. According to the assumed marked graph semantics, each actor consumes exactly one token from each input channel and produces one token on each output channel upon firing.

### B. MULTI-CAST ACTORS

Generally, an application graph might contain so-called *multi-cast actors* $a_\mathbf{m} \in A_\mathbf{M} \subset A$, e.g., actor $a_2 \in A_\mathbf{M}$ in Fig. 2a. In each actor firing of a multi-cast actor $a_\mathbf{m}$, one token is consumed from its *input channel* $c_{in}$, and for each *output channel* $c_{out}$, one token is produced containing copied data of the consumed token. To exemplify, actor $a_2$ copies each token consumed from input channel $c_1$ to actors $a_3$ and $a_4$ by producing for each output channel $c_{out} \in \{c_2, c_3\}$ a token containing identical data.

Formally, each multi-cast actor $a_\mathbf{m} \in A_\mathbf{M}$ has exactly one input channel and multiple output channels, as specified in Eq. (1). The size of the tokens contained in the input and output channels must be identical (see Eq. (2)), e.g., $\varphi(c_3) = \varphi(c_2) = \varphi(c_1)$. Finally, there must not be any initial tokens in the output channels, and the channel capacity of all output channels must be identical (see Eq. (3)), e.g., $\delta(c_2) = \delta(c_3) = 0$ and $\gamma(c_2) = \gamma(c_3)$.

$$\forall a_\mathbf{m} \in A_\mathbf{M} : |(C \times \{a_\mathbf{m}\}) \cap E| = 1 \wedge |(\{a_\mathbf{m}\} \times C) \cap E| \geq 1 \quad (1)$$
$$\wedge\, \forall (c_{in}, a_\mathbf{m}), (a_\mathbf{m}, c_{out}), (a_\mathbf{m}, c'_{out}) \in E : \varphi(c_{in}) = \varphi(c_{out}) \quad (2)$$
$$\wedge\, \delta(c_{out}) = 0 \wedge \gamma(c_{out}) = \gamma(c'_{out}) \quad (3)$$

Each multi-cast actor represents a memory footprint reduction opportunity by replacing it and its adjacent channels with an MRB, as shown and explained in the caption of Fig. 2.

### C. MULTI-READER BUFFER REALIZATION

A concept for unifying multiple FIFOs carrying identical data was first introduced by [12] (there called broadcast-FIFO). However, the proposed broadcast-FIFO has slightly altered semantics compared to the behavior of the multiple point-to-point FIFO channels it replaces. A concept preserving FIFO semantics called Multi-Reader Buffer (MRB) has then been presented in [9], which will be explained in the following.

By definition, an MRB $c_\mathbf{m}$ has one writer $a_\mathbf{w}$ and multiple readers $a_\mathbf{r} \in \{a \mid (c_\mathbf{m}, a) \in E\}$. For the example shown in Fig. 2b, the MRB $c_{\{1,2,3\}}$ has the writer $a_1$, and the actors $a_3$ and $a_4$ are its readers. An MRB has a write index $\omega_{c_\mathbf{m}} \in \{0, 1, \ldots, \gamma(c_\mathbf{m}) - 1\}$ that indicates the next position in $c_\mathbf{m}$'s buffer to be filled with the next token produced by the writer. Moreover, for each reader $a_\mathbf{r}$, there is a read index $\rho_{c_\mathbf{m}, a_\mathbf{r}} \in \{-1, 0, 1, \ldots, \gamma(c_\mathbf{m}) - 1\}$ that indicates the position in $c_\mathbf{m}$'s buffer from which the reader will consume the next token. The special value $-1$ denotes that $c_\mathbf{m}$ is empty from $a_\mathbf{r}$'s perspective.

Then, the number of available tokens $\mathsf{T}(c_\mathbf{m}, a_\mathbf{r})$ from the perspective of a reader $a_\mathbf{r}$ and the number of free places $\mathsf{F}(c_\mathbf{m})$ in $c_\mathbf{m}$ from the perspective of the writer $a_\mathbf{w}$ can be determined as follows:

$$\mathsf{T}(c_\mathbf{m}, a_\mathbf{r}) = \begin{cases} 0 & \text{if } \rho_{c_\mathbf{m}, a_\mathbf{r}} = -1 \\ ((\omega_{c_\mathbf{m}} - \rho_{c_\mathbf{m}, a_\mathbf{r}} - 1) \bmod \gamma(c_\mathbf{m})) + 1 & \text{otherwise} \end{cases}$$
$$\mathsf{F}(c_\mathbf{m}) = \gamma(c_\mathbf{m}) - \max_{(c_\mathbf{m}, a_\mathbf{r}) \in E} \mathsf{T}(c_\mathbf{m}, a_\mathbf{r})$$

It is worth noting that the presented MRB realization presented here can support even multi-rate dataflow. To demonstrate this, assume that the writer $a_\mathbf{w}$ produces $\psi(a_\mathbf{w})$ tokens and a reader $a_\mathbf{r}$ consumes $\kappa(a_\mathbf{r})$ tokens upon firing. Naturally, the writer $a_\mathbf{w}$ can only fire when $\mathsf{F}(c_\mathbf{m}) \geq \psi(a_\mathbf{w})$ holds. Similarly, $\mathsf{T}(c_\mathbf{m}, a_\mathbf{r}) \geq \kappa(a_\mathbf{r})$ must be satisfied for a reader $a_\mathbf{r}$ to fire.

When firing actor $a_\mathbf{w}$, each read index $\rho_{c_\mathbf{m}, a_\mathbf{r}}$ with value $-1$ (i.e., indicating that the MRB is empty from $a_\mathbf{r}$'s perspective) is set to the value $\omega_{c_\mathbf{m}}$ (Eq. (4)). Then, Eq. (5) is applied, which advances the writer index $\omega_{c_\mathbf{m}}$ by the number of produced tokens.

$$\forall_{(c_\mathbf{m}, a_\mathbf{r}) \in E} \; \rho_{c_\mathbf{m}, a_\mathbf{r}} \leftarrow \begin{cases} \omega_{c_\mathbf{m}} & \text{if } \rho_{c_\mathbf{m}, a_\mathbf{r}} = -1 \\ \rho_{c_\mathbf{m}, a_\mathbf{r}} & \text{otherwise} \end{cases} \quad (4)$$
$$\omega(c_\mathbf{m}) \leftarrow (\omega_{c_\mathbf{m}} + \psi(a_\mathbf{w})) \bmod \gamma(c_\mathbf{m}) \quad (5)$$

Accordingly, upon each firing of a reader $a_\mathbf{r}$, the corresponding read index $\rho_{c_\mathbf{m}, a_\mathbf{r}}$ is updated as follows:

$$\rho_{c_\mathbf{m}, a_\mathbf{r}} \leftarrow \begin{cases} -1 & \text{if } \mathsf{T}(c_\mathbf{m}, a_\mathbf{r}) = \kappa(a_\mathbf{r}) \\ (\rho_{c_\mathbf{m}, a_\mathbf{r}} + \kappa(a_\mathbf{r})) \bmod \gamma(c_\mathbf{m}) & \text{otherwise} \end{cases}$$

To exemplify, the MRB's read and write indices after various firings of the connected actors $a_1$, $a_3$, and $a_4$ are depicted in Fig. 3. There, actors $a_1$, $a_3$, and $a_4$ are, respectively, associated with the write index $\omega_{c_{\{1,2,3\}}}$ and the read indices $\rho_{c_{\{1,2,3\}}, a_3}$ and $\rho_{c_{\{1,2,3\}}, a_4}$.

Assuming the MRB is initially empty, these read and write indices have the values shown in Fig. 3a. Thus, $\mathsf{T}(c_{\{1,2,3\}}, a_3) = \mathsf{T}(c_{\{1,2,3\}}, a_4) = 0$ and $\mathsf{F}(c_{\{1,2,3\}}) = \gamma(c_{\{1,2,3\}}) - \max\{0, 0\} = 4$. At this point (see Fig. 3a), it is only possible to perform write operations. Before firing $a_1$, we must check if there is at least one free place available for the produced token, i.e., $\mathsf{F}(c_{\{1,2,3\}}) = 4 \geq 1$.

Next, assume actor $a_1$ fires three times, resulting in the state shown in Fig. 3b. There, the write index $\omega_{c_{\{1,2,3\}}}$ has advanced to 3, pointing to the next free place in the MRB's buffer. The

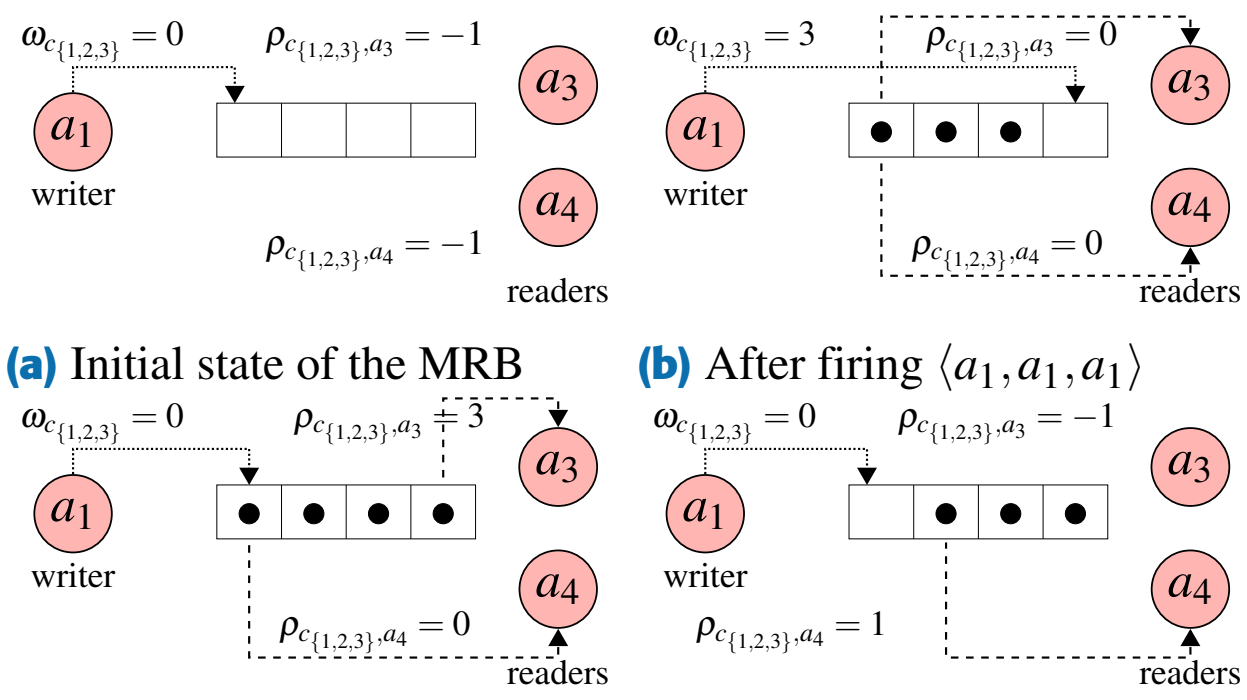

**(a)** Initial state of the MRB **(b)** After firing $\langle a_1, a_1, a_1 \rangle$ **(c)** MRB after $\langle a_3, a_3, a_3, a_1 \rangle$ **(d)** MRB after $\langle a_4, a_3 \rangle$ read indices $\rho_{c_{\{1,2,3\}},a_3}$ and $\rho_{c_{\{1,2,3\}},a_4}$ have been updated during the first firing of actor $a_1$ from $-1$ to 0, pointing to the first token contained in the MRB. At this point (see Fig. 3b), we can also perform read operations. Before firing a reader, we need to verify if there exist sufficient tokens to be consumed by the reader. For instance, we are able to fire actor $a_3$ because $\mathsf{T}(c_{\{1,2,3\}}, a_3) = ((3-0-1) \bmod 4) + 1 = 3 \geq 1$.

**FIGURE 3.** MRB with one write index (pointer) indicating the location of the next token to be written. Moreover, each reading actor requires an index pointing to the position of the next token to read.

After firing the sequence $\langle a_3, a_3, a_3, a_1 \rangle$, the resulting state is shown in Fig. 3c. There, the readers track different information about the state of the MRB. The reader $a_3$ points to $\rho_{c_{\{1,2,3\}},a_3} = 3$ and observes $\mathsf{T}(c_{\{1,2,3\}}, a_3) = ((0-3-1) \bmod 4) + 1 = 1$ token, whereas reader $a_4$ points to $\rho_{c_{\{1,2,3\}},a_4} = 0$ and observes $\mathsf{T}(c_{\{1,2,3\}}, a_4) = ((0-0-1) \bmod 4) + 1 = 4$ tokens. From the perspective of the writer $a_1$, the MRB is full. At this point (see Fig. 3c), let the firing sequence $\langle a_4, a_3 \rangle$ be observed.

The resulting state of the MRB is shown in Fig. 3d. From the perspective of $a_3$, the MRB is empty, i.e., $\rho_{c_{\{1,2,3\}},a_3}$ is $-1$. The token placed at position 0 has been consumed because $a_4$ has read it now, seeing $\mathsf{T}(c_{\{1,2,3\}}, a_4) = ((0-1-1) \bmod 4) + 1 = 3$ more tokens. From the perspective of $a_1$, there is one free place as $\mathsf{F}(c_{\{1,2,3\}}) = \gamma(c_{\{1,2,3\}}) - \max\{0,3\} = 4-3 = 1$.

### D. ARCHITECTURE GRAPH

A heterogeneous many-core target architecture, e.g., as depicted in the right part of Fig. 1, can be modeled formally by an abstract *architecture graph*:

*Definition 2.2 (Architecture Graph):* An architecture graph $g_R$ is a tuple $(R, L)$ composed of a set of vertices $R$ modeling hardware resources and a set of edges $L \subseteq R \times R$ denoting communication links between resources.

Here, the set of vertices $R = P \cup Q \cup H$ represents the resources of the architecture where each $p \in P$ denotes a core, each $q \in Q$ a memory, and each $h \in H$ an interconnect. The set of cores $P$ is partitioned into sets $P_{\vartheta_1}, P_{\vartheta_2}, \ldots P_{\vartheta_{|\Theta|}}$. Each set $P_\vartheta$ describes the set of cores of identical core type $\vartheta \in \Theta$.

The set of memory resources $Q = Q_P \cup Q_T \cup \{q_{global}\}$ can be partitioned into *core-local memories* ($q_{p_i} \in Q_P$), *tile-local memories* ($q_{T_j} \in Q_T$), and the *global memory* ($q_{global}$). Each core $p_i \in P$ has a core-local memory $q_{p_i}$ reachable via a link $(p_i, q_{p_i}) \in L$. Each memory $q \in Q$ has a *capacity* $W_q$, which denotes the number of bytes that can be stored in the memory.

The set of interconnects $H$ is partitioned into the Network-on-Chip (NoC) ($h_{NoC} \in H$) and a set of crossbars $h_T \in H_T = H \setminus \{h_{NoC}\}$. Each interconnect $h \in H$ is annotated with its bandwidth $B_h$, which is used to calculate data transfer delays. The time required to transport $\eta$ bytes of data over a crossbar $h_T$ can be calculated as $\eta / B_{h_T}$.

Resources of a given architecture, excluding the NoC and the global memory ($q_{global}$), i.e., processors, local memories, tile-local memories, and crossbars, are organized as a set of tiles $\mathbb{T}$. Each tile $T \in \mathbb{T}$ consists of a set of cores and their core-local memories, a tile-local memory, and a tile crossbar connecting the cores and memories of the tile. As each resource belongs to exactly one tile, tiles are (i) *disjoint*, i.e., $\forall T_i, T_j \in \mathbb{T} : T_i \cap T_j = \emptyset$ where $i \neq j$ and (ii) *covering*, i.e., $\cup_{T \in \mathbb{T}} T = R \setminus \{q_{global}, h_{NoC}\}$.

Intra-tile communication is provided by links connecting each core and memory of the respective tile via the tile crossbar. To exemplify, consider the tile $T_1$ presented in Fig. 1. It is composed of six cores $\{p_1, \ldots, p_6\}$, six core-local memories $\{q_{p_1}, \ldots, q_{p_6}\}$, the tile-local memory $q_{T_1}$, and the tile-crossbar $h_{T_1}$. Each core in tile $T_1$ has an exclusive communication link with its corresponding core-local memory, e.g., there exists a link $(p_i, q_{p_i})$ that connects core $p_i$ with its memory $q_{p_i}$. Moreover, each memory of the tile can be reached via the tile-crossbar $h_{T_1}$. If core $p_1$ sends data to $p_4$, such data will traverse the tile-crossbar $h_{T_1}$ via the links $(p_1, h_{T_1})$ and $(h_{T_1}, q_{p_4})$ to be stored in the core-local memory $q_{p_4}$ of core $p_4$.

For inter-tile communication, links are provided that connect each tile to the NoC ($h_{NoC}$), which in turn is connected to the global memory ($q_{global}$).

The set of resources involved in a data transfer between a core $p$ and a memory $q$ will be denoted by a *routing function* $\mathscr{R} : P \times Q \to \mathbb{P}(R)$[1], as explained in the following.

In the simplest case, a data transfer happens between a core $p_i$ and its local memory $q_{p_i}$. Then, no interconnect resources are involved, i.e., $\mathscr{R}(p_i, q_{p_i}) = \{p_i, q_{p_i}\}$.

Else, if the core $p$ and the memory $q$ share the same tile ($\exists T_j \in \mathbb{T} : p, q \in T_j$), an *intra-tile* data transfer is performed. In this case, the data transfer only traverses the tile crossbar $h_{T_j}$, i.e., $\mathscr{R}(p, q) = \{p, h_{T_j}, q\}$.

Otherwise, an *inter-tile* transfer is needed as the core $p$ and the memory $q$ are allocated in different tiles, i.e., $p \in T_j$, $q \in T_k$, and $T_j \neq T_k$. Then, the data needs to travel over the tile crossbar $h_{T_j}$ of the tile containing the core $p$, the NoC interconnect $h_{NoC}$, and the tile crossbar $h_{T_k}$ of the tile containing the memory $q$, i.e., $\mathscr{R}(p, q) = \{p, h_{T_j}, h_{NoC}, h_{T_k}, q\}$.

In all other cases, the global memory is used, and the involved interconnect resources are the tile crossbar $h_{T_j}$ and the NoC, i.e., $\mathscr{R}(p, q_{global}) = \{p, h_{T_j}, h_{NoC}, q_{global}\}$.

[1] Here, the $\mathbb{P}(X)$ notation denotes the power set of the set $X$, i.e., the set of all subsets of $X$.

### E. SPECIFICATION GRAPH

To perform explorations of allocations and mappings of actors to cores as well as channels to memories, a specification finally contains a set of *mapping edges* $M = M_A \cup M_C$ that is partitioned into a set of potential mappings $M_A = \{(a,p) \in A \times P \mid \exists \vartheta \in \Theta : p \in P_\vartheta \wedge \tau(a,\vartheta) \neq \perp\}^2$ of actors to cores and a set of potential mappings $M_C = C \times Q$ of channels to memories. These mapping edges specify that every memory can store each channel and that each actor $a$ can be mapped to every core $p \in P_\vartheta$ of a type $\vartheta$ that can execute the actor $a$. With these definitions, a *specification graph* can be defined as follows:

*Definition 2.3 (Specification Graph):* A specification graph $g_\mathrm{S}$ is a tuple $(V_\mathrm{S}, E_\mathrm{S})$ composed of a set of vertices $V_\mathrm{S}$ and a set of edges $E_\mathrm{S}$. The set of vertices $V_\mathrm{S} = A \cup C \cup R$ is formed from the union of vertices of the application graph $g_\mathrm{A}$ and the architecture graph $g_\mathrm{R}$. Similarly, the set of edges $E_\mathrm{S} = E \cup L \cup M$ is formed from the union of edges of both graphs and the set of *mapping edges*.

Figure 1 illustrates an example of an application graph, an architecture graph, and an exemplified set of actor-to-core and channel-to-memory mappings.

## III. DEFINITION OF THE DESIGN SPACE

This section introduces the *design space* of selective MRB replacements, formalizes the concept of actor and channel bindings, and illustrates the principles of actor and communication scheduling.

### A. SELECTIVE MRB REPLACEMENT

As discussed in Section II-B, each multi-cast actor represents an opportunity for memory footprint reduction by replacing it and its adjacent channels with an MRB, as shown in Fig. 2. However, replacing a multi-cast actor with an MRB may also lead to an increase in the *execution period* [9], which is defined as the time interval between two successive iterations of execution of a given application graph. Hence, which multi-cast actors are replaced by MRBs needs to be explored to trade between period and memory footprint, both to be minimized. For this purpose, we define a multi-cast actor *replacement* function $\xi : A_\mathbf{M} \to \{0,1\}$ to determine for each multi-cast actor $a_\mathbf{m} \in A_\mathbf{M}$ if it should be replaced by an MRB ($\xi(a_\mathbf{m}) = 1$) or kept ($\xi(a_\mathbf{m}) = 0$).

Formally, the replacement of selected multi-cast actors with MRBs for a given application graph $g_\mathrm{A}$ (e.g., as illustrated in Fig. 2a) can be realized by a graph transformation as detailed in Algorithm 1, leading to a transformed application graph $g_{\tilde{\mathrm{A}}}$ (e.g., as shown in Fig. 2b), where the selected multi-cast actors and the channels connected to them have been replaced by their corresponding MRBs.

### B. ACTOR AND CHANNEL BINDINGS

Next, determining an implementation of a transformed application graph $g_{\tilde{\mathrm{A}}}$ on an architecture requires a binding (i) of each actor to a processor, which is described by a set $\beta_A \subseteq M_A$ called *actor bindings*, and (ii) of each channel to a memory, which is described by a set $\beta_C \subseteq M_C$ called *channel bindings*. Moreover, each actor and channel must be bound to exactly one core (see Eq. (6)), respectively, memory (see Eq. (7)). Finally, the channels bound to a memory $q \in Q$ must not exceed its capacity (see Eq. (8)). A set of *feasible bindings* $\beta = \beta_A \cup \beta_C$ must satisfy Eqs. (6) to (8).

$$\forall a \in A : \qquad |\beta_A \cap (\{a\} \times P)| = 1 \tag{6}$$

$$\forall c \in C : \qquad |\beta_C \cap (\{c\} \times Q)| = 1 \tag{7}$$

$$\forall q \in Q : \qquad \sum_{(c,q) \in \beta_C} \gamma(c) \cdot \varphi(c) \leq W_q \tag{8}$$

The number of cores $\alpha(\vartheta)$ allocated of a given type $\vartheta$ can then be implicitly derived from the actor binding $\beta_A$ as *allocation* $\alpha$.

$$\alpha(\vartheta) = |\{p \in P_\vartheta \mid \exists (a,p) \in \beta_A\}| \tag{9}$$

For example, the actor and channel bindings shown in Fig. 4 are given by $\beta_A = \{(a_3,p_1),(a_4,p_2),(a_1,p_3),(a_5,p_3)\}$ and $\beta_C = \{(c_4,q_{p_1}),(c_5,q_{p_2}),(c_\mathbf{m},q_{p_3})\}$, respectively. From these, the core allocations $\alpha(\vartheta_1) = 2$, $\alpha(\vartheta_2) = 1$, and $\alpha(\vartheta_3) = 0$ can be derived. Moreover, the notation $\beta_A(a)$ and $\beta_C(c)$ denote the core and memory the actor $a$, respectively, channel $c$ are bound to, e.g., $\beta_A(a_3) = p_1$ and $\beta_C(c_4) = q_{p_1}$.

While, in principle, each channel $c \in C$ can be bound to any memory $q \in Q$, it makes sense to constrain the design space to be explored such that a channel will not be bound to a core-local memory of a core that does not at all access the channel data. Similarly, tile-local memories of tiles containing no core accessing the channel data can also be excluded. As a result, only five binding alternatives exist for each channel: (PROD) the core-local memory $q_{p_{prod}}$ of the core $p_{prod}$ producing the data, (TILE-PROD) the tile-local memory $q_{T_{prod}}$ of the tile $T_{prod}$ containing the core producing the data, (CONS) the core-local memory $q_{p_{cons}}$ of the core $p_{cons}$ consuming the data,

**Algorithm 1:** Selective MRB Replacement

**Function** substituteMRBs($g_A, \xi$)
  **Input :** Application graph $g_\mathrm{A}$ and function $\xi$
  **Output:** Transformed application graph $g_{\tilde{\mathrm{A}}}$
  $g_{\tilde{\mathrm{A}}} \leftarrow g_\mathrm{A}$ // Let $g_{\tilde{\mathrm{A}}}$ be a copy of $g_\mathrm{A}$
  /* Only replace $a_\mathbf{m}$ when $\xi(a_\mathbf{m}) = 1$ */
  **foreach** $a_\mathbf{m} \in A_\mathbf{M}$ *where* $\xi(a_\mathbf{m}) = 1$ **do**
    /* Let $C_\mathrm{del}$ be the set of all channels that are adjacent to $a_\mathbf{m}$ */
    $C_\mathrm{del} \leftarrow \{c \in g_{\tilde{\mathrm{A}}}.C \mid (c,a_\mathbf{m}) \in g_{\tilde{\mathrm{A}}}.E \vee (a_\mathbf{m},c) \in g_{\tilde{\mathrm{A}}}.E\}$
    /* New MRB channel $c_\mathbf{m}$ */
    $c_\mathbf{m} \leftarrow \mathrm{createMRB}(C_\mathrm{del})$
    /* Add an input edge to $c_\mathbf{m}$, replacing the ones for $a_\mathbf{m}$ */
    **foreach** $(a,c) \in g_{\tilde{\mathrm{A}}}.E$ *where* $c \in C_\mathrm{del} \wedge a \neq a_\mathbf{m}$ **do**
      $g_{\tilde{\mathrm{A}}}.E \leftarrow \{(a,c_\mathbf{m})\} \cup g_{\tilde{\mathrm{A}}}.E \setminus \{(a,c),(c,a_\mathbf{m})\}$
    /* Add output edges from $c_\mathbf{m}$, replacing the ones for $a_\mathbf{m}$ */
    **foreach** $(c,a) \in g_{\tilde{\mathrm{A}}}.E$ *where* $c \in C_\mathrm{del} \wedge a \neq a_\mathbf{m}$ **do**
      $g_{\tilde{\mathrm{A}}}.E \leftarrow \{(c_\mathbf{m},a)\} \cup g_{\tilde{\mathrm{A}}}.E \setminus \{(a_\mathbf{m},c),(c,a)\}$
    $g_{\tilde{\mathrm{A}}}.A \leftarrow g_{\tilde{\mathrm{A}}}.A \setminus \{a_\mathbf{m}\}$ // Remove $a_\mathbf{m}$
    $g_{\tilde{\mathrm{A}}}.C \leftarrow \{c_\mathbf{m}\} \cup g_{\tilde{\mathrm{A}}}.C \setminus C_\mathrm{del}$ // Replace $C_\mathrm{del}$ by $c_\mathbf{m}$
  **return** $g_{\tilde{\mathrm{A}}}$

[2] Remember, $\tau(a,\vartheta) = \perp$ denotes that an actor $a$ cannot be mapped to a particular core type $\vartheta$.

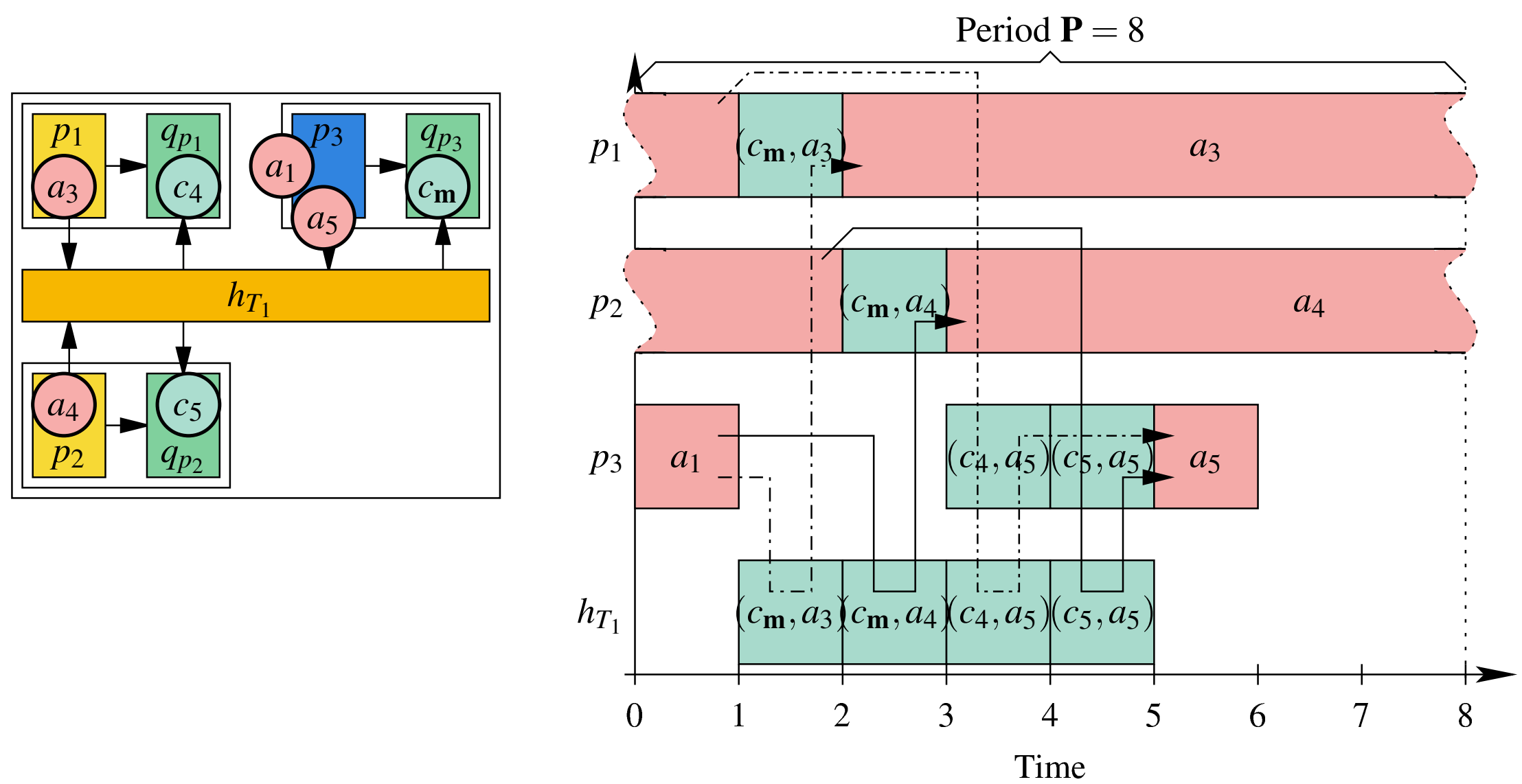

**FIGURE 4.** Example of a schedule with period $\mathbf{P} = 8$ time steps (right) for the transformed application graph $g_{\tilde{A}}$ from Fig. 2b where actors $a_1$ and $a_5$ are bound to core $p_3$, actor $a_3$ and channel $c_4$ are bound to core $p_1$ and its core-local memory $q_{p_1}$, as well as actor $a_4$ and channel $c_5$ are bound to core $p_2$ and its core-local memory $q_{p_2}$. For better visualization, we use $c_\mathbf{m}$ to refer to the MRB $c_{\{1,2,3\}}$, which is bound to the core-local memory $q_{p_3}$ of core $p_3$ (left). The light red boxes in the Gantt chart shown to the right denote actor executions, while the light green boxes denote read operations, e.g., the light green box containing $(c_4, a_5)$ denotes a read of a token contained in channel $c_4$ by the actor $a_5$. The data dependencies of the application graph $g_{\tilde{A}}$ are depicted by the solid and dotted dashed directed edges in the Gantt chart. For example, the solid directed edge from actor $a_1$ over the read communication $(c_\mathbf{m}, a_4)$ to actor $a_4$ represents the data dependency between actors $a_1$ and $a_4$ communicated via the MRB $c_\mathbf{m}$. The Gantt chart does not depict the corresponding write $(a_1, c_\mathbf{m})$ as the MRB $c_\mathbf{m}$ is bound to the core-local memory $q_{p_3}$ of the core $p_3$, where actor $a_1$ is bound to. Thus, the write communication is assumed to be part of the execution of actor $a_1$ itself.

(TILE-CONS) the tile-local memory $q_{T_{cons}}$ of the tile $T_{cons}$ containing the core consuming the data, or (GLOBAL) the global memory $q_{global}$.

In the following, these five options are represented by a *channel decision* function $C_\mathbf{d} : C \rightarrow \{$GLOBAL, TILE-PROD, TILE-CONS, PROD, CONS$\}$, which shall be explored rather than exploring channel bindings directly. Concrete channel bindings $\beta_C$ can then be determined via Algorithm 2 from the channel decisions, channel capacities, and actor bindings in such a way that Eqs. (7) and (8) are satisfied. Algorithm 2 determines for each channel $c \in C$ a concrete binding according to the channel decision $C_\mathbf{d}(c)$ in case memory capacities $W_q$ are not exceeded. Otherwise, a fallback solution is determined according to the case statements. It can be proven that a feasible binding is always found for each channel $c \in C$ by binding $c$ to the global memory $q_{global}$ that is assumed to be large enough to store all the buffer data related to the channels of a given application.

Algorithm 2 derives the channel bindings. For the running example in Fig. 4, we obtain $\beta_C = \{(c_4, q_{p_1}), (c_5, q_{p_2}), (c_\mathbf{m}, q_{p_3})\}$ from the channel decisions $C_\mathbf{d}(c_4) = C_\mathbf{d}(c_5) = C_\mathbf{d}(c_\mathbf{m}) =$ PROD and the actor bindings $\beta_A = \{(a_3, p_1), (a_4, p_2), (a_1, p_3), (a_5, p_3)\}$. Algorithm 2 thereby prefers to bind channels to core-local memories. If the core-local memory ($q_{p_3}$ in the running example) did not have a sufficient capacity to accommodate the MRB channel $c_\mathbf{m}$, Algorithm 2 would bind $c_\mathbf{m}$ to the tile-local memory $q_{T_1}$, and if even $q_{T_1}$ would also have an insufficient capacity, the channel $c_\mathbf{m}$ would finally be bound to the global memory $q_{global}$.

### C. PERIODIC SCHEDULING OF ACTORS AND COMMUNICATION

In the following, we consider the optimization and generation of static periodic schedules with an assumed uninterrupted execution of actors and communications. We also assume that each actor executes on the same core for each iteration of the dataflow graph. As it is assumed that the underlying DFG of a given application graph $g_{\tilde{A}}$ is a marked graph [3], for each actor $a \in g_{\tilde{A}}.A$, read $(c, a) \in g_{\tilde{A}}.E$, as well as write $(a, c) \in g_{\tilde{A}}.E$ operation, we need to determine exactly one *start time* $s_a$, $s_{(c,a)}$, and $s_{(a,c)}$, respectively, which repeats with the period $\mathbf{P}$. Thus, the actors and edges of the application graph $g_{\tilde{A}}$ together define the set of tasks to be scheduled, i.e., $\mathbf{t} \in \mathbf{T} = g_{\tilde{A}}.A \cup g_{\tilde{A}}.E$.

For example, consider the schedule with a period of $\mathbf{P} = 7$ depicted in Fig. 5 with actor start times as follows: $s_{a_1} = 0$, $s_{a_2} = 1$, $s_{a_3} = 3$, $s_{a_4} = 4$, and $s_{a_5} = 13$. Note that the start time of actor $a_5$ is greater than the period. Therefore, the firing of actor $a_5$ depicted in the schedule at time step 6 belongs to the previous iteration. Naturally, start times also need to be determined for the read and write operations, e.g., $s_{(a_2,c_2)} = 2$, $s_{(a_2,c_3)} = 3$, $s_{(c_4,a_5)} = 11$, and $s_{(c_5,a_5)} = 12$ for the

**Algorithm 2:** Determine Channel Bindings $\beta_C$

**Function** determineChannelBindings($C_{\mathbf{d}}, \gamma, \beta_A$)
  **Input** : Channel decision function $C_{\mathbf{d}}$, channel capacity function $\gamma$, and the set of actor bindings $\beta_A$
  **Output:** The set of channel bindings $\beta_C$
  $w_q \leftarrow 0 \ \forall q \in Q$ // Start memory usage $w_q$ from 0
  $\beta_C \leftarrow \emptyset$ // Start with empty bindings $\beta_C$
  // Derive binding for each channel $c \in C$
  **foreach** $c \in C$ **do**
    /* Derive $a_{prod}$, $p_{prod}$, $T_{prod}$, $a_{cons}$, $p_{cons}$, and $T_{cons}$ for channel $c$ */
    $a_{prod} \in A$ such that $(a_{prod}, c) \in E$ // Derive $a_{prod}$
    $p_{prod} \in P$ such that $(a_{prod}, p_{prod}) \in \beta_A$ // Derive $p_{prod}$
    $T_{prod} \in \mathbb{T}$ such that $p_{prod} \in T_{prod}$ // Derive $T_{prod}$
    $a_{cons} \in A$ such that $(c, a_{cons}) \in E$ // Derive $a_{cons}$
    $p_{cons} \in P$ such that $(a_{cons}, p_{cons}) \in \beta_A$ // Derive $p_{cons}$
    $T_{cons} \in \mathbb{T}$ such that $p_{cons} \in T_{cons}$ // Derive $T_{cons}$
    **switch** $C_{\mathbf{d}}(c)$ **do**
      **case** PROD **do**
        **if** $w_{q_{p_{prod}}} + \gamma(c) \cdot \varphi(c) \leq W_{q_{p_{prod}}}$ **then**
          // Bind $c$ to $q_{p_{prod}}$
          $\beta_C \leftarrow \beta_C \cup \{(c, q_{p_{prod}})\}$
          $w_{q_{p_{prod}}} \leftarrow w_{q_{p_{prod}}} + \gamma(c) \cdot \varphi(c)$
          **break**
        // $q_{p_{prod}}$ too small, try $q_{T_{prod}}$ next
      **case** TILE-PROD **do**
        **if** $w_{q_{T_{prod}}} + \gamma(c) \cdot \varphi(c) \leq W_{q_{T_{prod}}}$ **then**
          // Bind $c$ to $q_{T_{prod}}$
          $\beta_C \leftarrow \beta_C \cup \{(c, q_{T_{prod}})\}$
          $w_{q_{T_{prod}}} \leftarrow w_{q_{T_{prod}}} + \gamma(c) \cdot \varphi(c)$
          **break**
        // $q_{T_{prod}}$ too small, bind to $q_{global}$
        $\beta_C \leftarrow \beta_C \cup \{(c, q_{global})\}$
        **break**
      **case** CONS **do**
        **if** $w_{q_{p_{cons}}} + \gamma(c) \cdot \varphi(c) \leq W_{q_{p_{cons}}}$ **then**
          // Bind $c$ to $q_{p_{cons}}$
          $\beta_C \leftarrow \beta_C \cup \{(c, q_{p_{cons}})\}$
          $w_{q_{p_{cons}}} \leftarrow w_{q_{p_{cons}}} + \gamma(c) \cdot \varphi(c)$
          **break**
        // $q_{p_{cons}}$ too small, try $q_{T_{cons}}$ next
      **case** TILE-CONS **do**
        **if** $w_{q_{T_{cons}}} + \gamma(c) \cdot \varphi(c) \leq W_{q_{T_{cons}}}$ **then**
          // Bind $c$ to $q_{T_{cons}}$
          $\beta_C \leftarrow \beta_C \cup \{(c, q_{T_{cons}})\}$
          $w_{q_{T_{cons}}} \leftarrow w_{q_{T_{cons}}} + \gamma(c) \cdot \varphi(c)$
          **break**
        // $q_{T_{cons}}$ too small, bind to $q_{global}$
      **case** GLOBAL **do**
        // Bind $c$ to $q_{global}$
        $\beta_C \leftarrow \beta_C \cup \{(c, q_{global})\}$
  **return** $\beta_C$

write and read operations shown in the schedule. The read and write operations with assumed zero communication time (i.e., read and write operations not involving any interconnect resource), which are not depicted in the schedule, have the following start times: $s_{(a_1,c_1)} = s_{(c_1,a_2)} = 1$ (i.e., after actor $a_1$ has finished and before actor $a_2$ starts), $s_{(c_2,a_3)} = 3$ (i.e., before actor $a_3$ starts), $s_{(c_3,a_4)} = 4$ (i.e., before actor $a_4$ starts), $s_{(a_3,c_4)} = 10$ (i.e., after actor $a_3$ finishes), and $s_{(a_4,c_5)} = 11$ (i.e., after actor $a_4$ finishes).

Furthermore, for each actor $a$, its execution time is denoted by $\tau_a$, derivable from the actor bindings $\beta_A$ as follows:

$$\tau_a = \tau(a, \vartheta) \text{ where } \vartheta \in \Theta \text{ such that } \beta_A(a) \in P_\vartheta \qquad (10)$$

For example, the actor execution times $\tau_{a_1} = \tau_{a_2} = \tau_{a_5} = 1$ and $\tau_{a_3} = \tau_{a_4} = 7$ correspond to those depicted in the schedule shown in Fig. 5.

The time required for one token to be read from, respectively, written to channel $c$ by actor $a$ is denoted by $\tau_{(c,a)}$ and $\tau_{(a,c)}$. In the following, these times are derived from the token size $\varphi(c)$ and the interconnect bandwidth $B_h$ of the interconnect $h$ with the minimal bandwidth that is traversed by the communication:

$$\tau_{(c,a)} = \tau_{(a,c)} = \varphi(c) / \min_{h \in \mathscr{R}(\beta_A(a), \beta_C(c)) \cap H} B_h \qquad (11)$$

As a consequence, read and write operations that do not traverse at least one interconnect resource have zero communication time, e.g., $\tau_{(a_1,c_1)} = \tau_{(c_1,a_2)} = \tau_{(c_2,a_3)} = \tau_{(a_3,c_4)} = \tau_{(c_3,a_4)} = \tau_{(a_4,c_5)} = 0$ for the actor and channel bindings given in Fig. 5. Such communication operations directly access a core-local memory $q_{p_i}$ from the corresponding core $p_i$. In this case, the communication is assumed to be part of the execution of the actor performing the read or write operation. In other cases, the traversed interconnect resource $h$ with an assumed minimal bandwidth $B_h$ leads to a non-zero communication time. In the example above, $\tau_{(a_2,c_2)} = \tau_{(a_2,c_3)} = \tau_{(c_4,a_5)} = \tau_{(c_5,a_5)} = 1$, as visualized in the schedule in Fig. 5.

Finally, let $A_r$ and $\mathbf{T}_r$ denote the set of actors, respectively, tasks mapped to a resource $r$. Formally, $A_r$ can be derived from the set of actor bindings $\beta_A$ as follows:

$$A_r = \{a \in g_{\tilde{\mathrm{A}}}.A \mid r = \beta_A(a)\} \qquad (12)$$

For a fully formal definition of $\mathbf{T}_r$, we extend the domain of the routing function $\mathscr{R}$ to also contain all edges $e \in g_{\tilde{\mathrm{A}}}.E$ of the application graph $g_{\tilde{\mathrm{A}}}$. Given the bindings $\beta_A$ and $\beta_C$, let the set of resources involved by a write operation $e = (a,c)$ or read operation $e = (c,a)$ be denoted by $\mathscr{R}(a,c) = \mathscr{R}(\beta_A(a), \beta_C(c))$, respectively, $\mathscr{R}(c,a) = \mathscr{R}(\beta_A(a), \beta_C(c))$. With this extension, $\mathbf{T}_r$ is given by:

$$\mathbf{T}_r = \{e \in g_{\tilde{\mathrm{A}}}.E \mid r \in \mathscr{R}(e)\} \cup A_r \qquad (13)$$

For example, the set of all actors bound to core $p_3$, as shown in Fig. 5, is given by $A_{p_3} = \{a_1, a_2, a_5\}$. Including read and write operations executed by core $p_3$ results in the set $\mathbf{T}_{p_3} = A_{p_3} \cup \{(a_1,c_1), (c_1,a_2), (a_2,c_2), (a_2,c_3), (c_4,a_5), (c_5,a_5)\}$. Note that the write $(a_1,c_1)$ and the read $(c_1,a_2)$ are not shown in the schedule depicted in Fig. 5, as these are assumed to have zero communication times. Moreover, read and write operations are, in general, bound to multiple resources, as they are bound to the core where the data is produced or consumed as well as all traversed interconnect resources, e.g., the read and write operations $(a_2,c_2)$, $(a_2,c_3)$, $(c_4,a_5)$, and $(c_5,a_5)$ are not only executed by core $p_3$ but are also traversing the interconnect $h_{T_1}$, i.e., $\mathbf{T}_{h_{T_1}} = \{(a_2,c_2), (a_2,c_3), (c_4,a_5), (c_5,a_5)\}$.

### D. TRADE-OFFS BETWEEN THE MINIMIZATION OF MEMORY FOOTPRINT AND THE ACHIEVABLE PERIOD

Replacing a multi-cast actor and its adjacent channels with an MRB has as its primary purpose the reduction of the memory footprint (see Fig. 2). Moreover, this transformation removes both the need to execute the multi-cast actor and its communication. Nonetheless, there are cases where an MRB replacement is detrimental to (i.e., it increases) the execution period $\mathbf{P}$. To illustrate this, Figs. 4 and 5 present two periodic schedules obtained from the specification shown in Fig. 1. One can see that the schedule shown in Fig. 4 utilizing an MRB has a longer period, i.e., $\mathbf{P} = 8$, than the schedule with period $\mathbf{P} = 7$ depicted in Fig. 5, where the multi-cast actor $a_2$ has been retained. The timings in Figs. 4 and 5 are chosen for illustrative purposes to demonstrate the impact of MRBs and the existing trade-off in the specification. In both schedules, the same actor-to-core binding is assumed for actors $a_1, a_3, a_4$, and $a_5$, i.e., actors $a_1$ and $a_5$ are bound to core $p_3$, actor $a_3$ is bound to core $p_1$, and actor $a_4$ is bound to core $p_2$. Moreover, channels $c_4$ and $c_5$ are bound to the core-local memories $q_{p_1}$ and $q_{p_2}$, respectively.

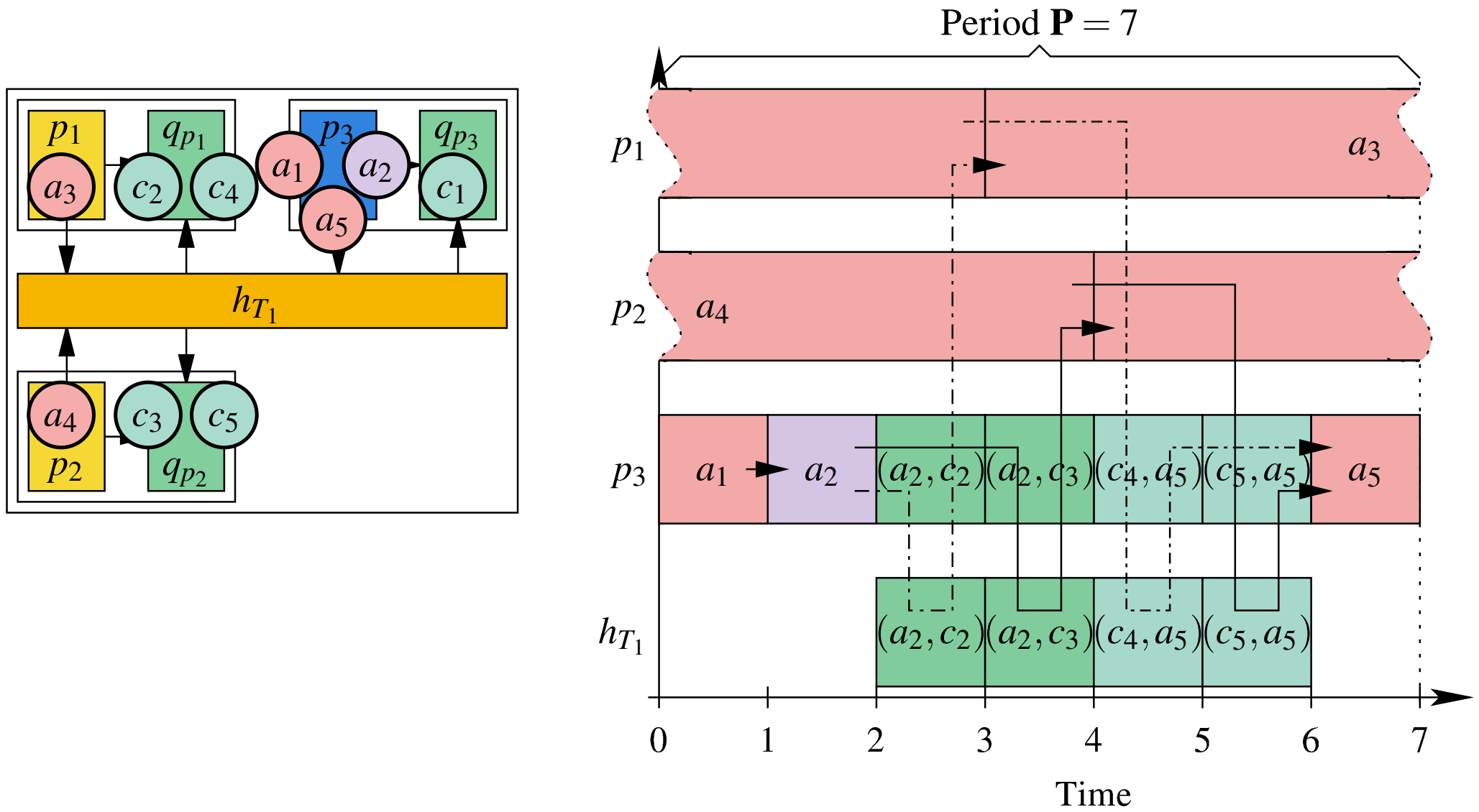

**FIGURE 5.** Schedule with a period of $\mathbf{P} = 7$ (shown to the right) for the application graph $g_A$ from Fig. 2a. Actor $a_3$ is bound to core $p_1$, actor $a_4$ is bound to core $p_2$, and actors $a_1$, $a_2$, and $a_5$ are bound to core $p_3$. Channels $c_2$ and $c_4$ are bound to the core-local memory $q_{p_1}$, channels $c_3$ and $c_5$ are bound to core-local memory $q_{p_2}$, and channel $c_1$ is bound to core-local memory $q_{p_3}$ (shown to the left). The light red boxes and the light violet box (for the multi-cast actor $a_2$) in the Gantt chart shown to the right denote actor executions, the green boxes represent write operations, and the light green boxes indicate read operations. To exemplify, the green box containing $(a_2, c_2)$ represents a write of a token to channel $c_2$ by the actor $a_2$, and the light green box containing $(c_4, a_5)$ indicates a read of a token contained in channel $c_4$ by the actor $a_5$. Similarly to Fig. 4, the data dependencies of the application graph $g_A$ are depicted by the solid and dotted dashed directed edges in the Gantt chart. The read and write from and to channel $c_1$ are not shown in the Gantt chart as both the write $(a_1, c_1)$ of actor $a_1$ to channel $c_1$ and the read $(c_1, a_2)$ of actor $a_2$ from channel $c_1$ access the core-local memory $q_{p_3}$ of the core $p_3$ that executes both actors $a_1$ and $a_2$. Thus, the corresponding write and read communication times are zero, i.e., $\tau_{(a_1,c_1)} = \tau_{(c_1,a_2)} = 0$, as the communication is assumed to be part of the execution of the actors themselves. The same situation holds for the read from channel $c_2$ and write to channel $c_4$ by actor $a_3$ as well as the read from channel $c_3$ and write to channel $c_5$ by actor $a_4$, i.e., $\tau_{(c_2,a_3)} = \tau_{(a_3,c_4)} = \tau_{(c_3,a_4)} = \tau_{(a_4,c_5)} = 0$.

As mentioned previously, the illustrated schedules are distinguished whether they employ an MRB or retain the multi-cast actor $a_2$. To exemplify, in Fig. 4, the MRB $c_\mathbf{m}$ mapped to the core-local memory $q_{p_3}$ replaces the multi-cast actor $a_2$ and its connected channels $c_1$, $c_2$, and $c_3$. Thus, both actors $a_3$ and $a_4$ have to read from memory $q_{p_3}$ (i.e., the reads $(c_\mathbf{m}, a_3)$ and $(c_\mathbf{m}, a_4)$), resulting in an additional delay of 1 time unit, increasing the period to $\mathbf{P} = 8$. Moreover, binding the MRB $c_\mathbf{m}$ to either core-local memory $q_{p_1}$ or core-local memory $q_{p_2}$ does not improve the situation as, respectively, actor $a_4$ or actor $a_3$ has to perform a read, delaying its execution by 1 time unit. For the example, the only way to obtain a schedule with a period of $\mathbf{P} = 7$ is to have copies of the output data of actor $a_1$ in both core-local memories $q_{p_1}$ and $q_{p_2}$ (e.g., as shown in Fig. 5), but this is the exact situation that is prevented when employing an MRB, as MRBs are used to avoid any data duplication. Hence, no schedule with a period of $\mathbf{P} = 7$ exists when an MRB replaces the multi-cast actor $a_2$.

In contrast, the schedule depicted in Fig. 5 retains the multi-cast actor $a_2$ (bound to core $p_3$) and its connected channels $c_1$, $c_2$, and $c_3$. Channel $c_1$ is bound to core-local memory $q_{p_3}$, while channels $c_2$ and $c_3$ are bound to core-local memories $q_{p_1}$ and $q_{p_2}$, respectively. Thus, the input data needed to fire actors $a_3$ and $a_4$ are already contained in the core-local memories (i.e., $q_{p_1}$ and $q_{p_2}$) of the cores the actors are bound to (i.e., $p_1$ and $p_2$). Moreover, their output channels ($c_4$ and $c_5$) are also bound to these core-local memories. Thus, the core $p_1$ can execute actor $a_3$ without any read or write overhead. The same holds for core $p_2$ and its bound actor $a_4$. Instead, the communication overhead to move the input and output data of actors $a_3$ and $a_4$ to and from the core-local memories $q_{p_1}$ and $q_{p_2}$ is spent by core $p_3$, which was previously under-utilized in the schedule depicted in Fig. 4. Core $p_3$ executes the multi-cast actor $a_2$, which provides the input data of actors $a_3$ and $a_4$ via the writes $(a_2, c_2)$ and $(a_2, c_3)$, and the actor $a_5$ (also bound to core $p_3$) is fetching the output data of actors $a_3$ and $a_4$ via the reads $(c_4, a_5)$ and $(c_5, a_5)$. This enables a schedule of period $\mathbf{P} = 7$, as the cores $p_1$ and $p_2$ are no longer burdened with any communication overhead.

Moreover, this also demonstrates that the channel decisions must be explored to obtain this optimal period of $\mathbf{P} = 7$. Otherwise, in case of a fixed channel decision, the actors $a_3$ and $a_4$ would need to execute a communication operation, e.g., a read operation when the data stays at the producer (PROD) or a write operation when the data has to be moved to the consumer (CONS). Only with the channel decisions $C_{\mathbf{d}}(c_2) = C_{\mathbf{d}}(c_3) = \texttt{CONS}$ and $C_{\mathbf{d}}(c_4) = C_{\mathbf{d}}(c_5) = \texttt{PROD}$ is a schedule with a period of $\mathbf{P} = 7$ possible.

In summary, replacing every multi-cast actor with an MRB enables minimal memory footprint implementations, but this may create an impact on the minimal achievable period. Thus, minimal period implementations require both optimization of the actor and channel bindings as well as a selective decision for each multi-cast actor on whether or not to perform MRB replacement. In the following, we present our design space exploration approach to minimize the execution period, memory footprint, and core cost.

## IV. DESIGN SPACE EXPLORATION

Allocation of resources, binding, and scheduling a DFG onto a heterogeneous many-core system is a Multi-objective Optimization Problem (MOP) [5, 10], and trade-offs exist and shall be explored between different objectives, e.g., execution period, memory footprint, and core cost. In general, there is no single best solution but a set of Pareto-optimal solutions that trade the different objectives against each other.

Moreover, the introduced design space of bindings and schedules is huge even for small applications and a modest number of processors, memories, and communication resources, such as the example shown in Fig. 1. Thus, finding the actual set of Pareto-optimal solutions is an intractable problem that can only be approximated via heuristics. For this purpose, many state-of-the-art Electronic System Level (ESL) design flows employ *meta-heuristic optimization* techniques based on MOEAs [5, 7, 15]. The advantage of such population-based techniques is that the search space is sampled in parallel and that not only one compromise solution but an approximation of the Pareto-front is found after several generations of offspring as a result of the DSE. However, whereas MOEAs have been shown to provide quite good results for allocation and binding problems [5, 10], it is difficult to find good encodings for feasible schedules of operations.

Indeed, pure meta-heuristic optimization techniques, while applicable to a broad domain of problems, are often too generic. This general applicability can be traded for a better optimization performance, e.g., quality of found solutions or required runtime to obtain these solutions, by employing problem-specific heuristics. Hence, it is beneficial to *integrate problem knowledge* into meta-heuristic optimization techniques – restricting their general applicability to a particular domain but improving optimization performance.

In this paper, we propose a new hybrid DSE approach in which the exploration of the design space is split between (i) a MOEA to explore the space of *multi-cast actor replacement function* $\xi$ (encoded as a binary string), *channel decision function* $C_{\mathbf{d}}$ (integer encoding), and the *set of actor bindings* $\beta_A$ (integer encoding). To find a schedule minimizing the execution period $\mathbf{P}$ for a given solution candidate, (ii) a specialized scheduling algorithm is applied. This so-called hybrid *decoding* process is illustrated in Fig. 6.

For decoding, we first propose an exact formulation for the related scheduling problem and subsequently introduce our heuristic CAPS-HMS. The ILP-based decoding will obtain a schedule with minimal period for a given set of actor bindings and channel decisions but may suffer from long evaluation times. In contrast, our heuristic CAPS-HMS will allow for a much faster evaluation of solution candidates but does not guarantee to find the exact minimal period.

In both alternative approaches, Algorithm 1 is applied first to compute a transformed application graph $g_{\tilde{\mathrm{A}}}$ containing the MRBs decided by the DSE via the multi-cast actor replacement function $\xi$. This function $\xi$, the channel decision function $C_{\mathbf{d}}$, and the set of actor bindings $\beta_A$ together form the genotype $\mathscr{G}$. In both cases, the genotype will be decoded into the *phenotype* representing the period $\mathbf{P}$, the set of actor and channel bindings $\beta$, and the channel capacity function $\gamma$. Based on the phenotype, the evaluators finally determine the quality of the solution candidate under evaluation with respect to the design *objectives*. For our mapping and scheduling problem, the objectives are the minimization of (i) the execution period $\mathbf{P}$, (ii) the memory footprint $\mathbf{M_F} = \sum_{c \in g_{\tilde{\mathrm{A}}}.C} \gamma(c) \cdot \varphi(c)$, and (iii) the core cost $\mathbf{K} = \sum_{\vartheta \in \Theta} \alpha(\vartheta) \cdot \mathbf{K}_{\vartheta}$.[3]

## V. DECODING

In the following, we present and later evaluate two decoding approaches: (i) an integer linear program and (ii) a novel periodic scheduling heuristic called CAPS-HMS for heterogeneous multi-core platforms with hierarchical memory organizations and integrating the scheduling of actors and communications between actors.

### A. ILP-BASED DECODING

First, we explain our ILP-based decoding approach, as shown in Algorithm 3. This algorithm decodes the genotype $\mathscr{G}$ into the corresponding phenotype $(\mathbf{P}, \beta, \gamma)$, as shown in Fig. 6.

Note that the scheduling via ILP is performed in a loop (Lines 2 to 6). The reason is that for an ILP-derived schedule, the channel capacities might need to be increased to execute this schedule (Line 5), and the channel bindings might need to be modified in consequence to accommodate the enlarged channels (Line 3). The loop terminates when all channels fit into the memories they are bound to (Line 6).

[3] Here, $\mathbf{K}_{\vartheta}$ denotes the cost of a core of type $\vartheta$.

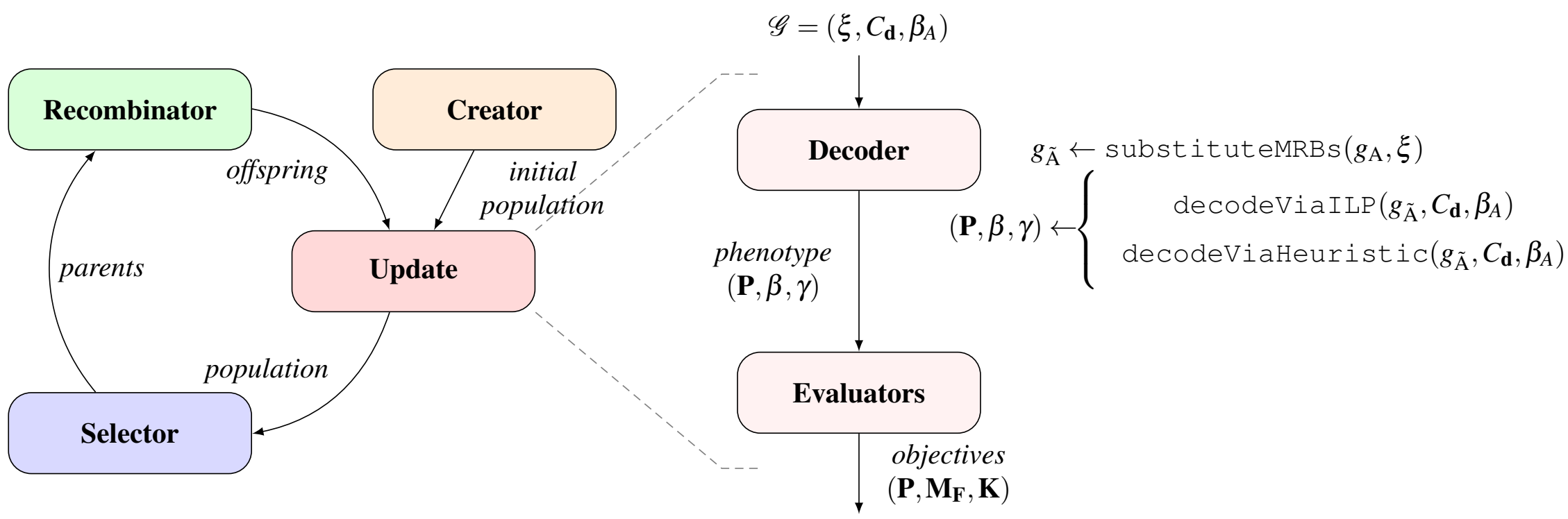

**FIGURE 6.** Overview of our hybrid DSE approach using MOEAs. The instance *creator* generates random *genotypes* for an *initial population* that forms the starting point of the iterative optimization process. A genotype $\mathscr{G}$ is the genetic representation of a solution candidate. For each (new) solution candidate, *update* executes a user-defined decoder and then applies evaluator functions on this candidate, i.e., either `decodeViaILP` or `decodeViaHeuristic`, depending on whether the ILP-based or heuristic-based approach is used. The decoder transforms the genotype into the *phenotype* representing the solution candidate's characteristics of interest, e.g., the period $\mathbf{P}$, the bindings $\beta$, and the channel capacities $\gamma$. Based on the phenotype, the evaluators determine the quality of the solution candidate under evaluation with respect to the design *objectives*, e.g., period $\mathbf{P}$, memory footprint $\mathbf{M_F}$, and core cost $\mathbf{K}$. From the resulting population, the *selector* chooses *parents* with superior solution quality. Finally, the recombinator generate *offsprings* by recombining and mutating the genotype of the selected parents. Our approach has been realized using the DSE framework *OpenDSE* [13] and its underlying MOEA-based optimization framework Opt4J [14].

**Algorithm 3:** ILP-based Decoding

**Function** `decodeViaILP` $(g_{\tilde{A}}, C_{\mathbf{d}}, \beta_A)$
 **Input :** Application graph $g_{\tilde{A}}$, channel decision function $C_{\mathbf{d}}$, and the set of actor bindings $\beta_A$
 **Output:** Period $\mathbf{P}$, set of bindings $\beta$, and the channel capacity function $\gamma$
 **do**
  $\beta_C \leftarrow$ `determineChannelBindings`$(C_{\mathbf{d}}, \gamma, \beta_A)$
  Solve ILP

$$\text{minimize} \quad \mathbf{P} \in \mathbb{R}_0^+ \tag{14}$$
$$\forall \mathbf{t} \in \mathbf{T} \quad s_{\mathbf{t}} \in \mathbb{R}_0^+ \tag{15}$$
$$\forall (a,c),(c,a') \in g_{\tilde{A}}.E \quad s_{(a,c)} + \tau_{(a,c)} - \mathbf{P} \cdot \delta(c) \leq s_{(c,a')} \tag{16}$$
$$\forall (c,a) \in g_{\tilde{A}}.E \quad s_{(c,a)} + \tau_{(c,a)} \leq s_a \tag{17}$$
$$\forall (a,c) \in g_{\tilde{A}}.E \quad s_a + \tau_a \leq s_{(a,c)} \tag{18}$$
$$\forall r \in R \; \forall \mathbf{t},\mathbf{t}' \in \mathbf{T}_r \quad s_{\mathbf{t}} + \tau_{\mathbf{t}} - \mathbf{P} \leq s_{\mathbf{t}'} \tag{19}$$
$$\forall r \in H \cup P \; \forall \mathbf{t},\mathbf{t}' \in \mathbf{T}_r \wedge \mathbf{t} \neq \mathbf{t}' \quad e_{\mathbf{t},\mathbf{t}'} \in \{0,1\} \tag{20}$$
$$e_{\mathbf{t},\mathbf{t}'} + e_{\mathbf{t}',\mathbf{t}} = 1 \tag{21}$$
$$\forall h \in H \; \forall \mathbf{t},\mathbf{t}' \in \mathbf{T}_h \wedge \mathbf{t} \neq \mathbf{t}' \quad s_{\mathbf{t}} + \tau_{\mathbf{t}} - D \cdot (1 - e_{\mathbf{t},\mathbf{t}'}) \leq s_{\mathbf{t}'} \tag{22}$$
$$\forall p \in P \; \forall a,a' \in A_p \wedge a \neq a' \; \forall \mathbf{t} \in OUT(a), \mathbf{t}' \in IN(a') \quad s_{\mathbf{t}} + \tau_{\mathbf{t}} - D \cdot (1 - e_{a,a'}) \leq s_{\mathbf{t}'} \tag{23}$$

  Increase $\gamma(c)$ to accommodate ILP schedule $\quad \forall c \in g_{\tilde{A}}.C$
 **while** $\exists q \in Q : \sum_{(c,q) \in \beta_C} \gamma(c) \cdot \varphi(c) > W_q$ ;
 **return** $(\mathbf{P}, \beta_A \cup \beta_C, \gamma)$

The objective of the ILP itself is the minimization of the execution period $\mathbf{P}$ (Eq. (14)). Moreover, for each task $\mathbf{t} \in \mathbf{T}$, the ILP determines a start time $s_{\mathbf{t}}$ (Eq. (15)). Equations (16) to (18) encode the data dependencies of the application graph $g_{\tilde{A}}$. In particular, Eq. (16) denotes that a token cannot be read from a channel $c$ before it has been written into it, also considering the number of initial tokens $\delta(c)$ of the channel. Equation (17) ensures that each actor can only start after all its reads from ingoing edges have been performed, and Eq. (18) enforces that each actor write can only start after its actor computation has finished. Equation (19) guarantees for each resource $r$ that all tasks $\mathbf{t} \in \mathbf{T}_r$ mapped to this resource are executed within a time interval of duration $\mathbf{P}$. Finally, to ensure a feasible schedule, the ILP must enforce that tasks mapped to the same resource have non-overlapping executions. For this purpose, sequentialization binary variables $e_{\mathbf{t},\mathbf{t}'}$ are introduced for each pair of tasks that share a resource (Eq. (20)). Here, $e_{\mathbf{t},\mathbf{t}'} = 1$ denotes that task $\mathbf{t}$ must finish before task $\mathbf{t}'$ is started. Thus, exactly either $e_{\mathbf{t},\mathbf{t}'}$ or $e_{\mathbf{t}',\mathbf{t}}$ must be one (Eq. (21)). These variables are then used to sequentialize the communication over the interconnects (Eq. (22)) and the actor executions performed by the cores (Eq. (23)). In these equations, $D \gg \mathbf{P}$ is a value much greater than the execution period, that is used to disable the sequentialization constraint that task $\mathbf{t}$ must finish before task $\mathbf{t}'$ is started in the case that $e_{\mathbf{t},\mathbf{t}'} = 0$. The sequentialization of actors mapped to the same core (Eq. (23)) is enforced indirectly by constraining that all write tasks $\mathbf{t} \in OUT(a)$ of actor $a$ are finished before the read tasks $\mathbf{t}' \in IN(a')$ of actor $a'$ are started. This ensures that all reads of an actor, then the actor itself, and finally, all writes of the actor are executed in sequence without interspersing of reads and writes of other actors into this sequence.

However, if actor $a$ is a sink actor (i.e., has no output edges) or actor $a'$ is a source actor (i.e., has no input edges), a simple definition of $OUT(a)$ and $IN(a')$ as the set of all output edges of actor $a$, respectively, the set of all input edges of actor $a'$ would fail to enforce the sequentialization that actor $a$ is completed before actor $a'$ fires. To handle these cases, $OUT(a)$ returns the set containing only the actor $a$ itself when this actor is a sink. Conversely, $IN(a')$ returns the set containing only the actor $a'$ itself when this actor is a source. Formally, $OUT(a)$ and $IN(a')$ are defined as follows:

$$OUT(a) = \begin{cases} E_{\mathrm{O}}(a) & \text{if } E_{\mathrm{O}}(a) \neq \emptyset \\ \{a\} & \text{otherwise} \end{cases} \qquad IN(a') = \begin{cases} E_{\mathrm{I}}(a') & \text{if } E_{\mathrm{I}}(a') \neq \emptyset \\ \{a'\} & \text{otherwise} \end{cases}$$

**Algorithm 4:** Heuristic-based Decoding

1 **Function** `decodeViaHeuristic`($g_{\tilde{\text{A}}}, C_{\mathbf{d}}, \beta_A$)
  **Input** : Application graph $g_{\tilde{\text{A}}}$, channel decision function $C_{\mathbf{d}}$, and the set of actor bindings $\beta_A$
  **Output:** Period $\mathbf{P}$, set of bindings $\beta$, and the channel capacity function $\gamma$
2   $\beta_C \leftarrow$ `determineChannelBindings`($C_{\mathbf{d}}, \gamma, \beta_A$)
3   $\mathbf{P} \leftarrow \max_{\forall r \in P \cup H} \left\lceil \sum_{\forall \mathbf{t} \in \mathbf{T}_r} \tau_{\mathbf{t}} \right\rceil$
4   **while true do**
5     **while** $\neg$CAPS-HMS($g_{\tilde{\text{A}}}, \beta_A, \beta_C, \mathbf{P}$) **do**
6       $\mathbf{P} \leftarrow \mathbf{P} + 1$
7     Increase $\gamma(c)$ to accommodate schedule $\forall c \in g_{\tilde{\text{A}}}.C$
8     **if** $\forall q \in Q : \sum_{(c,q) \in \beta_C} \gamma(c) \cdot \varphi(c) \leq W_q$ **then**
9       **break**
10     $\beta_C \leftarrow$ `determineChannelBindings`($C_{\mathbf{d}}, \gamma, \beta_A$)
11   **return** ($\mathbf{P}, \beta_A \cup \beta_C, \gamma$)

Here, $E_{\text{O}}(a) = \{(\hat{a}, \hat{c}) \in g_{\tilde{\text{A}}}.E_{\text{O}} \mid \hat{a} = a\}$ denotes the set of all output edges (i.e., write operations) of actor $a$ and, correspondingly, $E_{\text{I}}(a') = \{(\hat{c}, \hat{a}) \in g_{\tilde{\text{A}}}.E_{\text{I}} \mid \hat{a} = a'\}$ denotes the set of all input edges (i.e., read operations) of actor $a'$.

### B. HEURISTIC-BASED DECODING

To speed up evaluation during exploration, we propose an alternative heuristic-based decoding outlined in Algorithm 4. This algorithm decodes the input genotype $\mathscr{G}$ into the corresponding phenotype $(\mathbf{P}, \beta, \gamma)$, as shown in Fig. 6.

First, we determine an initial set of channel bindings $\beta_C$ in Line 2. Note that channels may need to be remapped later on (Line 10) if it turns out that channel capacities need to be increased (Line 7) to accommodate the found schedule and at least one channel no longer fits into the memory it is bound to (checked in Line 8). After initial channel bindings have been determined in Line 2, a lower bound for the period $\mathbf{P}$ is derived in Line 3 from the resource utilization of cores and interconnects. Consider Fig. 7 as an example, where bindings and timings are chosen for illustrative purposes with a communication time of one for all reads and writes, i.e., $\tau_{\mathbf{t}} = 1 \; \forall \mathbf{t} \in E$. The bottleneck resource in this example is the crossbar $h_{T_1}$ involved in five reads and five writes, leading to a lower bound of 10 for the period $\mathbf{P}$.

A concrete schedule is calculated by the proposed scheduler Communication-Aware Periodic Scheduling on Heterogeneous Many-core Systems (CAPS-HMS) depicted in Algorithm 5. CAPS-HMS is called with an application $g_{\tilde{\text{A}}}$, actor and channels bindings $\beta_A$ and $\beta_C$, and a candidate period $\mathbf{P}$. If a schedule with period $\mathbf{P}$ is found, **true** is returned, **false** otherwise. This is used by the loop in Lines 5 to 6 of Algorithm 4 to successively increase the period until a schedule is found. As discussed previously, channel capacities may need to be enlarged to accommodate the found schedule, possibly resulting in a need to remap channels no longer fitting into memory, necessitating a rescheduling with the updated channel bindings, as is done by the while loop in Lines 4 to 10. Otherwise, as soon as a schedule with a feasible period $\mathbf{P}$ is found and all channels fit into the memory they are bound to, Line 9 terminates the loop. Then, the resulting phenotype $(\mathbf{P}, \beta, \gamma)$ is returned in Line 11.

CAPS-HMS shown in Algorithm 5 follows a greedy strategy, where tasks are scheduled as soon as possible on the resources they are bound to. All tasks are assigned a start time of execution within a given interval $[0, \mathbf{P}[$, i.e., from 0 (included) to $\mathbf{P}$ (excluded). Ultimately, this interval will contain tasks from different iterations to optimize resource utilization. To obtain a schedule within the interval $[0, \mathbf{P}[$, CAPS-HMS schedules one iteration of the application graph $g_{\tilde{\text{A}}}$, thereby wrapping task executions finishing later than the period $\mathbf{P}$ back into the *schedule interval* $[0, \mathbf{P}[$ through modulo $\mathbf{P}$ computation. Assuming the task $\mathbf{t}$ is executed in the interval $[s_{\mathbf{t}}, s_{\mathbf{t}} + \tau_{\mathbf{t}}[$, then in the schedule interval $[0, \mathbf{P}[$, it will occupy the time region given by $f_{\text{wrap}}(\mathbf{P}, s_{\mathbf{t}}, \tau_{\mathbf{t}}) = \{t \bmod \mathbf{P} \mid s_{\mathbf{t}} \leq t < s_{\mathbf{t}} + \tau_{\mathbf{t}}\}$. For example, the execution of actor $a_3$ in the schedule depicted in Fig. 7 (to the right) is from 8 to 11, but it is wrapped into the schedule interval $[0, 10[$ with $f_{\text{wrap}}(10, 8, 3) = [8, 10[ \cup [0, 1[$.

During scheduling, the *resource utilization* of each core or interconnect resource $r \in R \setminus Q$ is tracked by a corresponding utilization set $\mathtt{U}_r \subseteq [0, \mathbf{P}[$ that contains all time intervals already occupied with scheduled tasks. Initially, all resources are free, i.e., the utilization sets are assigned the empty set (Line 2 in Algorithm 5). For example, in the state depicted by the partial schedule shown in the middle of Fig. 7, the actors $a_1$, $a_2$, and $a_3$ and all their read and write operations have already been scheduled. In this state, the heuristic is trying to schedule actor $a_4$ with its read and write operations, observing the utilization sets $\mathtt{U}_{h_{T_1}} = [1, 4[ \cup [5, 8[$, $\mathtt{U}_{p_1} = [0, 7[$, $\mathtt{U}_{p_2} = [0, 2[ \cup [7, 10[$, $\mathtt{U}_{p_3} = \mathtt{U}_{p_4} = \emptyset$.

The goal of the scheduling heuristic CAPS-HMS is to assign for each task $\mathbf{t} \in \mathbf{T}$ a as early as possible start time $s_{\mathbf{t}}$ that conforms with the given bindings and satisfies the data dependencies. Channel capacities are not considered during scheduling but are adjusted in Algorithm 4 to accommodate the found schedule. The start times are initialized with zero at algorithm start (Line 3 in Algorithm 5) as, later on, the heuristic only delays start times to conform to data dependencies and resource constraints. CAPS-HMS considers for each actor a priority given by the topological sorting of $g_{\tilde{\text{A}}}$ (see Line 4). During scheduling, the heuristic keeps track of actors to be scheduled with the list $\mathtt{L}$ of ready actors, which is initialized in Line 5 as all actors that are initially ready to be fired, e.g., because they are source actors or there is at least one initial token contained in all input channels of the actor. Before any actor is selected, the ready list $\mathtt{L}$ must be sorted in descending order using the previously assigned priority.

Actor scheduling is performed by the loop in Lines 6 to 24, which either *succeeds* (Line 25) when there are no longer any actors to be scheduled, i.e., $\mathtt{L} = \emptyset$, or *fails* (Line 24) when an actor can not be scheduled within the schedule interval $[0, \mathbf{P}[$ due to insufficient free time remaining on at least one resource to schedule the actor and its read and write operations. This failure is indicated by the error flag $\varpi$ checked in (Line 23). Within the scheduling loop, an actor $a$ to be scheduled is selected from the ready list $\mathtt{L}$, and its core $p$ onto which it is bound is derived from the bindings $\beta_A$ (Line 8). Then, the time $\tau'_a$ that an actor $a$, including its communication tasks, requires to be scheduled on core $p$ is computed. For this purpose, we sum the actor's execution time $\tau_a$ and the required times $\tau_{E_\mathrm{I}(a)}$ and $\tau_{E_\mathrm{O}(a)}$ for the actors read $\mathbf{t} \in E_\mathrm{I}(a)$ and write $\mathbf{t} \in E_\mathrm{O}(a)$ communication tasks (see Line 9). To exemplify, for the actor $a_1$, $\tau_{E_\mathrm{I}(a_1)} = 0$ as the actor is a source actor having no inputs, $\tau_{E_\mathrm{O}(a_1)} = 1$ as there is a single write task with communication time $\tau_{(a_1,c_1)} = 1$, resulting in an overall time including communication of $\tau'_{a_1} = \tau_{E_\mathrm{I}(a_1)} + \tau_{a_1} + \tau_{E_\mathrm{O}(a_1)} = 0+2+1 = 3$.

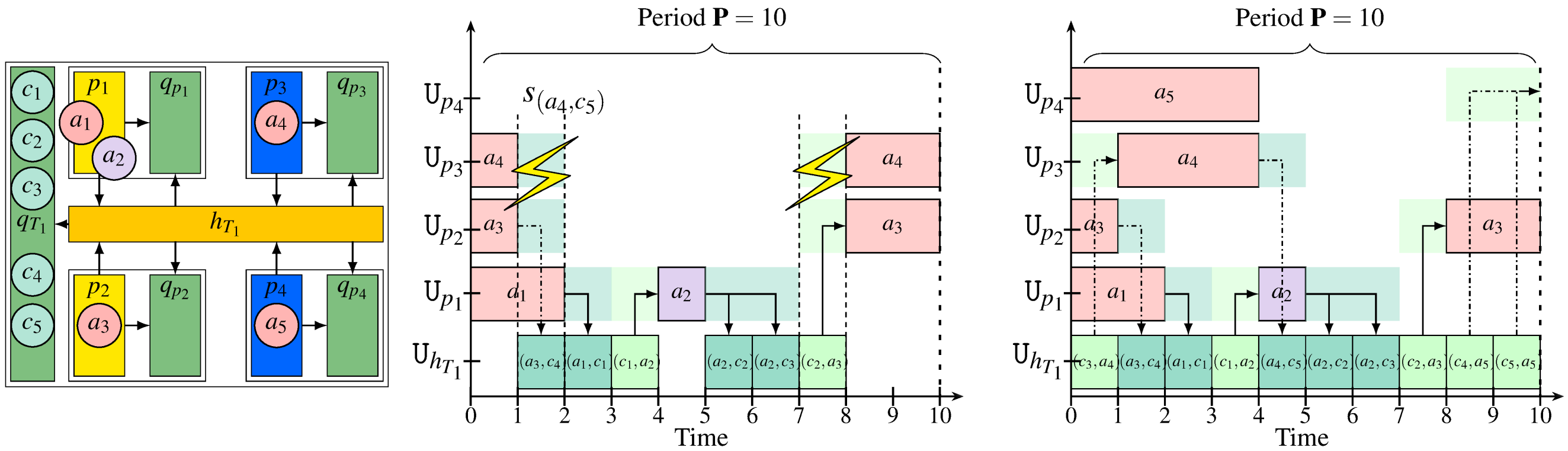

**FIGURE 7.** Left: Example of given actor and channel bindings, with all communication channels bound to $q_{T_1}$, actors $a_1$ and $a_2$ bound to core $p_1$, and actors $a_3$, $a_4$, and $a_5$ bound to cores $p_2$, $p_3$, and $p_4$, respectively. During scheduling, time intervals in the utilization sets $\mathtt{U}_r$ are allocated to task executions, e.g., in the partial schedule depicted in the middle, the execution of actor $a_1$ is allocating the time interval $[0,2[\subseteq \mathtt{U}_{p_1}$ in the utilization set of core $p_1$. This partial schedule represents the state when the actors $a_1$, $a_2$, and $a_3$ and all their read and write operations have already been scheduled, and the heuristic is trying to schedule actor $a_4$ with its read and write operations. In this state, the read $(c_3,a_4)$, execute (actor $a_4$), and write $(a_4,c_5)$ sequence has to be delayed from 7 to 10 due to contention in the interconnect $h_{T_1}$, i.e., the interconnect is already occupied by the read $(c_2,a_3)$ and the write $(a_3,c_4)$. The final schedule is depicted on the right, where all actors and their read and write operations are shown in the schedule interval $[0,10[$. Note that the depicted tasks are from different iterations, e.g., the execution of actor $a_4$ is from one iteration in the past, while the execution of actor $a_5$ is from two iterations in the past.

Next, the scheduling heuristics searches for a free time span $[s'_a, s'_a + \tau'_a[$ in the utilization set $\mathtt{U}_p$ of core $p$ (Line 11) to schedule the actor $a$ and its read and write operations. First, however, the error indicator $\varpi$ is set to **true** (Line 10). Later, in case the loop in Lines 11 to 22 successfully schedules the actor and its communication, the error flag $\varpi$ will be reset to **false** to indicate success, and the loop will be terminated (Line 22). In detail, Line 11 iterates over start times $s'_a$ that satisfies the data dependencies of actor $a$, i.e., $s'_a \geq s_a$, while simultaneously not overlapping with already scheduled tasks on core $p$, i.e., $\mathtt{U}_p \cap f_\mathrm{wrap}(\mathbf{P}, s'_a, \tau'_a) = \emptyset$. Moreover, as the schedule is $\mathbf{P}$ periodic, the heuristic only needs to examine start times $s'_a < s_a + \mathbf{P}$.

Next, the read $\mathbf{t}_{cns,i}$ and write $\mathbf{t}_{prd,i}$ operations of actor $a$ are identified in Lines 12 and 13. Then, for each read $\mathbf{t}_{cns,i}$ operation, a start time $s_{\mathbf{t}_{cns,i}}$ is assigned from the time span $[s'_a, s'_a + \tau_{E_\mathrm{I}(a)}[$ before the execution of actor $a$ (see Line 14). Conversely, for each write task $\mathbf{t}_{prd,i}$, a start time $s_{\mathbf{t}_{prd,i}}$ is assigned from the time span $[s'_a + \tau_{E_\mathrm{I}(a)} + \tau_a, s'_a + \tau'_a[$ just after the execution of actor $a$ (see Line 15). Subsequently, CAPS-HMS checks for each communication task $\mathbf{t} \in E_\mathrm{I}(a) \cup E_\mathrm{O}(a)$ if all the (interconnect) resources $r \in \mathscr{R}(\mathbf{t})$ traversed by the read or write operation $\mathbf{t}$ are available during the time span $[s_\mathbf{t}, s_\mathbf{t} + \tau_\mathbf{t}[$ in which the communication operation is performed (Line 16). If the interconnects are free, the

**Algorithm 5:** CAPS-HMS

```
1  Function CAPS-HMS (g_Ã, β_A, β_C, P)
     Input  : Application g_Ã, bindings β_A and β_C, and a candidate period P
     Output: true if a schedule for g_Ã of period P is found, false otherwise
2    U_r ← ∅   ∀r ∈ R \ Q
3    s_t ← 0   ∀t ∈ T = g_Ã.A ∪ g_Ã.E
4    z_a ← assign topological sorting [16] as priority ∀a ∈ g_Ã.A
5    L ← {a ∈ g_Ã.A | a is ready to fire }
6    while L ≠ ∅ do
7      Sort L in descending order using priority z
8      a ← f_Pop(L) ; p ← β_A(a)
9      τ_{E_I(a)} ← Σ_{t∈E_I(a)} τ_t ; τ_{E_O(a)} ← Σ_{t∈E_O(a)} τ_t ; τ'_a ← τ_{E_I(a)} + τ_a + τ_{E_O(a)}
10     ϖ ← true
11     foreach s'_a ∈ [s_a, s_a + P[ : U_p ∩ f_wrap(P, s'_a, τ'_a) = ∅ do
12       (t_{cns,1}, t_{cns,2}, ... t_{cns,|E_I(a)|}) ← E_I(a)
13       (t_{prd,1}, t_{prd,2}, ... t_{prd,|E_O(a)|}) ← E_O(a)
14       s_{t_{cns,i}} ← s'_a + Σ_{j=1}^{i-1} τ_{t_{cns,j}}   ∀i ∈ {1,...,|E_I(a)|}
15       s_{t_{prd,i}} ← s'_a + τ_{E_I(a)} + τ_a + Σ_{j=1}^{i-1} τ_{t_{prd,j}}   ∀i ∈ {1,...,|E_O(a)|}
16       if ∀t ∈ E_I(a) ∪ E_O(a), r ∈ ℛ(t) : U_r ∩ f_wrap(P, s_t, τ_t) = ∅ then
17         s_a ← s'_a + τ_{E_I(a)}
18         U_p ← U_p ∪ f_wrap(P, s_a, τ_a)
19         U_r ← U_r ∪ f_wrap(P, s_t, τ_t)   ∀t ∈ E_I(a) ∪ E_O(a), r ∈ ℛ(t)
20         s_{a'} ← max(s_{a'}, s'_a + τ'_a)
             ∀(a,c),(c,a') ∈ g_Ã.E ∧ δ(c) = 0
21         L ← (L \ {a}) ∪ {a' ∈ g_Ã.A | unscheduled actor a' is
             ready due to the firing of actor a}
22         ϖ ← false ; break
23     if ϖ then
24       return false
25   return true
```

start time of actor $a$ is updated (see Line 17), actor $a$ is scheduled to core $p$ (see Line 18), and each communication task $\mathbf{t} \in E_\mathrm{I}(a) \cup E_\mathrm{O}(a)$ is scheduled to its traversed resources $r \in \mathscr{R}(\mathbf{t})$ (see Line 19). Then, for each successor actor $a'$ of actor $a$, its start time $s_{a'}$ is updated to respect the data dependency between actor $a$ and $a'$ (Line 20). Next, actor $a$ is removed from the ready list $\mathtt{L}$, and all unscheduled actors $a'$ that have been enabled by firing actor $a$ are added to the ready list (Line 21). Finally, the foreach loop is terminated in Line 22 to continue scheduling the next actor until all the actors have been scheduled (Line 25) or there is insufficient free time remaining on at least one resource to schedule all actors and their read and write operations (Line 24).

We will see in Section VI that although our heuristic scheduler CAPS-HMS does not guarantee to determine a schedule of minimal period $\mathbf{P}$ for a given combination of graph, channel decision function, and actor bindings, it turns out to require much less execution time than using the ILP scheduling approach presented in Section V-A. When comparing related Pareto front qualities, we will also show that the degradation is little for many test applications. Particularly for large applications and complexity of the target architecture, the ILP solution times can become prohibitively long.

## VI. RESULTS

In the following, we conduct a series of different DSE experiments as shown in Fig. 6 to assess the effectiveness of our proposed ILP and CAPS-HMS heuristic in generating high-quality implementations when mapping dataflow applications onto the heterogeneous many-cores shown in Fig. 1. For each exploration, we employed the OpenDSE [13] framework using the NGSA-II elitist genetic algorithm [17] with a population size of 100 individuals, each generation generating 25 new individuals and the crossover rate set to 0.95. To measure the effects of selectively introducing MRBs, we implemented and compared three different exploration strategies: *Reference*, $MRB_{\text{Always}}$, and $MRB_{\text{Explore}}$. The genotype for the *Reference* strategy is $\mathscr{G} = (C_{\mathbf{d}}, \beta_A)$. The multi-cast actor replacement function $\xi$ is the all-zeros function. Thus, no multi-cast actor is replaced (i.e., $g_{\tilde{A}} = g_A$). In contrast, $MRB_{\text{Always}}$ also uses the genotype $\mathscr{G} = (C_{\mathbf{d}}, \beta_A)$ but assumes the all-ones function for $\xi$. Thus, each multi-cast actor is replaced by its corresponding MRB. Finally, strategy $MRB_{\text{Explore}}$ selectively explores for each multi-cast actor the choice of its replacement by an MRB by using the complete genotype $\mathscr{G} = (\xi, C_{\mathbf{d}}, \beta_A)$. Here, the binary string $\xi$ is determined during the optimization loop (see Fig. 6).

Orthogonal to the replacement of multi-cast actors by MRBs, we also decide on decoding the genotype of each implementation. Here, we observe the effects of decoding via an ILP (see Section V-A) or using CAPS-HMS (see Section V-B). Both return a phenotype $(\mathbf{P}, \mathbf{M_F}, \mathbf{K})$ composed of a minimum period to modulo schedule, the memory footprint, and the cost of cores of an implementation. Such a phenotype is used to evaluate the quality of each implementation. In the following, the combinations of strategy and way to decode a solution candidate result in six approaches. The approaches named $Reference^{ILP}$, $MRB^{ILP}_{\text{Always}}$, and $MRB^{ILP}_{\text{Explore}}$ explore the effects of introducing MRBs when each genotype is decoded using the ILP-based decoder. Conversely, the approaches named $Reference^{\text{CAPS-HMS}}$, $MRB^{\text{CAPS-HMS}}_{\text{Always}}$, and $MRB^{\text{CAPS-HMS}}_{\text{Explore}}$ use CAPS-HMS to decode the genotype.

The architecture used for our experiments (shown in Fig. 1) contains 24 cores organized into four tiles: $T_1, T_2, T_3$, and $T_4$. Inter-tile communication is supported via a network-on-chip $h_{NoC}$. A global memory $q_{global}$ provides off-chip storage. Internally, each tile comprises six cores connected to its correspondent local memory. Each core is of one of three core types: $\vartheta_1, \vartheta_2$, or $\vartheta_3$. For our experiments, the respective relative core costs have been chosen as $\mathbf{K}_{\vartheta_1} = 1.5$, $\mathbf{K}_{\vartheta_2} = 1.0$, or $\mathbf{K}_{\vartheta_3} = 0.5$. Faster cores are usually more expensive than slower ones. Thus, the slowest processors in the architecture are those of type $\vartheta_3$, and the fastest processors in the architecture are those of type $\vartheta_1$. The relative core costs thus approximately correlate to the speedup between the cores of different types, i.e., cores of type $\vartheta_1$ are $3\times$ faster than cores of type $\vartheta_3$, and cores of type $\vartheta_2$ are $2\times$ faster than cores of type $\vartheta_3$. Moreover, each tile supports intra-tile communication via a crossbar $h_T$ and a tile-local memory $q_T$. To observe the effects of the approaches under observation in a realistic environment, we constrain the size of each memory and the bandwidth of each interconnect resource. Accordingly, the core-local and tile-local memories can store up to 2.5 MiB and 50 MiB, respectively. We assume the global memory to be large enough to store all channels of the explored applications. Last, the bandwidth of each crossbar is 8 GiB/s, and the NoC bandwidth is 4 GiB/s.

We assume that each actor in the application can be mapped to any core in the architecture, and each channel might potentially be mapped to any memory. The optimization loop of the DSE explores the actor-to-core bindings $\beta_A$, whereas channel-to-memory bindings $\beta_C$ are then determined using Algorithm 2 (see Section III-B).

As discussed, the objectives to be minimized are the execution period $\mathbf{P}$ (see Section V), the memory footprint $\mathbf{M_F}$, and the cost $\mathbf{K}$ of allocated cores. We quantify the memory footprint of each application $g_{\tilde{A}}$ after decoding as follows:

$$\mathbf{M_F} = \sum_{c \in g_{\tilde{A}}.C} \gamma(c) \cdot \varphi(c) \tag{24}$$

This corresponds to the addition of the product of the token size ($\varphi$) and the adjusted channel capacity ($\gamma$) of each channel.

We calculate the core cost $\mathbf{K}$ of each implementation after decoding as given below:

$$\mathbf{K} = \sum_{\vartheta \in \Theta} \alpha(\vartheta) \cdot \mathbf{K}_{\vartheta} \tag{25}$$

As target applications, Table 1 presents a benchmark composed of three real-world image processing applications obtained from self-developed Matlab/Simulink test cases [6]. Shown in the table are also the number of actors, the number of channels, and the number of multi-cast actors contained in each application. Table 1 also shows for each application two memory footprints, $\mathbf{M_F}$ and $\mathbf{M_F}_{\text{min}}$, with the following semantics: $\mathbf{M_F}$ represents the minimal memory footprint of each application when all multi-cast actors are retained, while $\mathbf{M_F}_{\text{min}}$ represents the minimal memory footprint when each multi-cast actor is replaced by a corresponding MRB. To calculate both memory footprints $\mathbf{M_F}$ and $\mathbf{M_F}_{\text{min}}$, we use Eq. (24) and assume a channel capacity of exactly one token for all channels, i.e., $\forall c \in C : \gamma(c) = 1$.

Finally, as our applications are all acyclic, they are transformed in such a way that there is at least one initial token per channel, i.e., $\forall c \in C: \delta(c) \geq 1$, allowing lower execution periods to be reached.

### A. QUALITY OF FOUND IMPLEMENTATIONS

A Multi-objective Optimization Problem (MOP) generally does not have a single optimal solution due to the conflicting objectives. Instead, there exists a set of Pareto-optimal solutions. The set of all such solutions is known as the Pareto-front. As discussed previously, finding the actual Pareto front of the MOP considered in this paper is an intractable problem that can only be approximated. To obtain a good approximation of the Pareto front for each application, the Pareto-fronts found by all exploration runs for a given application utilizing all six considered combinations of exploration and decoding strategy are combined into a reference Pareto-front. This reference Pareto-front $S_{\text{Ref}}$ can be seen as the closest approximation of the actual Pareto-front achieved. The quality of each approach for each application can then be evaluated by comparing the Pareto-front approximations found by the five DSE runs performed for this application and approach combination against the application's reference Pareto-front. To facilitate such a comparison, quality measures are required for Pareto-front approximations that condense characteristics such as proximity to the reference Pareto front (the closer, the better) and diversity into a single measure [18]. For this purpose, we use the hypervolume [19] quality measure and normalize the reference Pareto-front $S_{\text{Ref}}$ and each Pareto-front $S$ found by an approach to only contain objective values between zero and one, i.e., $S_{\text{Ref}}, S \subset [0,1]^d$ where the number of objectives is given by $d = 3$. This normalization ensures that each objective is weighted equally in the hypervolume quality measure.

Then, given a (normalized) Pareto-front $S \subset [0,1]^d$, the hypervolume of $S$ is the measure of the region weakly dominated[4] by $S$ and bounded above by the reference point $\mathbf{1}$.

$$\text{hypervolume}(S) = \Lambda(\{q \in [0,1]^d \mid \exists p \in S : p \leq q\}) \quad (26)$$

There, $\Lambda(\cdot)$ denotes the Lebesgue measure [20]. The greater the hypervolume score is, the better a Pareto-front approximation $S$ is considered to be.

For each considered application and approach under investigation, five independent DSE runs were performed. To make the comparison of the approaches feasible and fair, each DSE run was given a maximum number of 2,500 generations, which is sufficient for all approaches to reach stagnation, i.e., no or very little further progress could be observed if the exploration runs longer. In each generation of the DSE, the set of non-dominated solutions[5] found so far is recorded. Thus, for a given application, approach, and generation $i$, there exists a set $\mathbf{S}^{\leq i}$ containing exactly five sets $S^{\leq i}$ of non-dominated solutions found until generation $i$, one for each DSE run. To evaluate the quality of each approach for each application, we average over the five DSE runs as follows:

$$\frac{1}{|\mathbf{S}^{\leq i}|} \cdot \sum_{S^{\leq i} \in \mathbf{S}^{\leq i}} \frac{\text{hypervolume}(S^{\leq i})}{\text{hypervolume}(S_{\text{Ref}})} \quad (27)$$

Fig. 8 presents for each explored application and approach the averaged relative hypervolume score, as defined by Eq. (27). There, the approaches implementing *Reference*, $MRB_{\text{Always}}$ and $MRB_{\text{Explore}}$ correspond to dashed, dashed-dotted and solid traces, respectively. Moreover, we distinguish approaches using the ILP decoder ($Reference^{ILP}$, $MRB^{ILP}_{\text{Always}}$ and $MRB^{ILP}_{\text{Explore}}$) and approaches using CAPS-HMS ($Reference^{\text{CAPS-HMS}}$, $MRB^{\text{CAPS-HMS}}_{\text{Always}}$ and $MRB^{\text{CAPS-HMS}}_{\text{Explore}}$) colored in red and blue, respectively. In the following, we discuss the obtained results.

**Key Observations:** First, we confirm our expectation that the replacement of multi-cast actors by MRBs results in better solutions according to the design objectives. The results presented in Figs. 8 and 9 show that regardless of the chosen decoding approach, either ILP (see solid red lines) or the CAPS-HMS heuristic (see solid blue lines), the selective exploration of MRB replacements performed by the $MRB_{\text{Explore}}$ strategy delivers better quality solutions in terms of the hypervolume score compared to the respective *Reference* approach. These improvements range from 28 % for the small Sobel application to 90 % for the large multicamera application.

Next, it can be observed that the $MRB_{\text{Explore}}$ strategy gains superiority to the $MRB_{\text{Always}}$ strategy for applications with a rising number of multi-cast actors. For example, for the Sobel application containing only 1 multi-cast actor, the hypervolume score is almost identical, but for the $\text{Sobel}_4$ application containing 4 multi-cast actors, the $MRB^{ILP}_{\text{Explore}}$ approach improves upon the $MRB^{ILP}_{\text{Always}}$ approach by 6 %. For the large multicamera application with 23 multi-cast actors, the improvement of $MRB^{\text{CAPS-HMS}}_{\text{Explore}}$ compared to $MRB^{\text{CAPS-HMS}}_{\text{Always}}$ is even 20 %.

Finally, it can be observed that the ILP-based decoder is superior to the CAPS-HMS heuristic for small to mid-sized applications, i.e., Sobel and $\text{Sobel}_4$, where utilizing the $MRB^{\text{CAPS-HMS}}_{\text{Explore}}$ approach is only slightly inferior by 7 %, respectively, 5 % in terms of the hypervolume score compared to the $MRB^{ILP}_{\text{Explore}}$ approach. In contrast, the $MRB^{\text{CAPS-HMS}}_{\text{Explore}}$ approach is superior for the large multicamera application by 67 %. This observation can be explained by the fact that the ILP-solver was provided with a timing budget of three seconds for each ILP-decoding to stay within a limit of 1 day for the overall exploration time for each DSE run. For the small Sobel application, the ILP-solver is almost always able to deliver a minimal period schedule for a given binding of actors and channels to cores and memories within the timing budget. This was less frequently possible for the larger $\text{Sobel}_4$ application, with feasible, but not necessarily optimal solutions returned by the ILP-solver. But for the multicamera application, the number of variables to be solved by the ILP drastically increases compared to the other smaller applications. There, the ILP-based decoder may only infrequently find an optimal solution within the time budget. If the ILP-solver finds no optimal solution within three seconds, it often delivered at least a feasible modulo-schedule $\mathbf{P}$ solution, thus explaining the inferiority over CAPS-HMS for medium and large problem sizes.

[4] A point $p \in \mathbb{R}^d$ *weakly dominates* a point $q \in \mathbb{R}^d$ if $p_i \leq q_i$ for all $1 \leq i \leq d$.

[5] In our context, the set of non-dominated solutions is an approximation of the Pareto-front of the three-objective optimization problem with period $\mathbf{P}$, the memory footprint $\mathbf{M_F}$, and core cost $\mathbf{K}$ to be minimized.

**TABLE 1.** Applications investigated during DSE runs. $\mathbf{M_F}$ corresponds to the minimal memory footprint in case all multi-cast actors are retained, while $\mathbf{M_{F\min}}$ denotes the case when each multi-cast actor is replaced by a corresponding MRB.

| Application | $\lvert A\rvert$ | $\lvert C\rvert$ | $\lvert A_{\mathbf{M}}\rvert$ | $\mathbf{M_F}$ [MiB] | $\mathbf{M_{F\min}}$ [MiB] |
|---|---|---|---|---|---|
| Sobel | 7 | 7 | 1 | 71.15 | 55.33 |
| $\text{Sobel}_4$ | 23 | 29 | 4 | 71.22 | 55.38 |
| Multicamera | 62 | 111 | 23 | 50.47 | 32.15 |

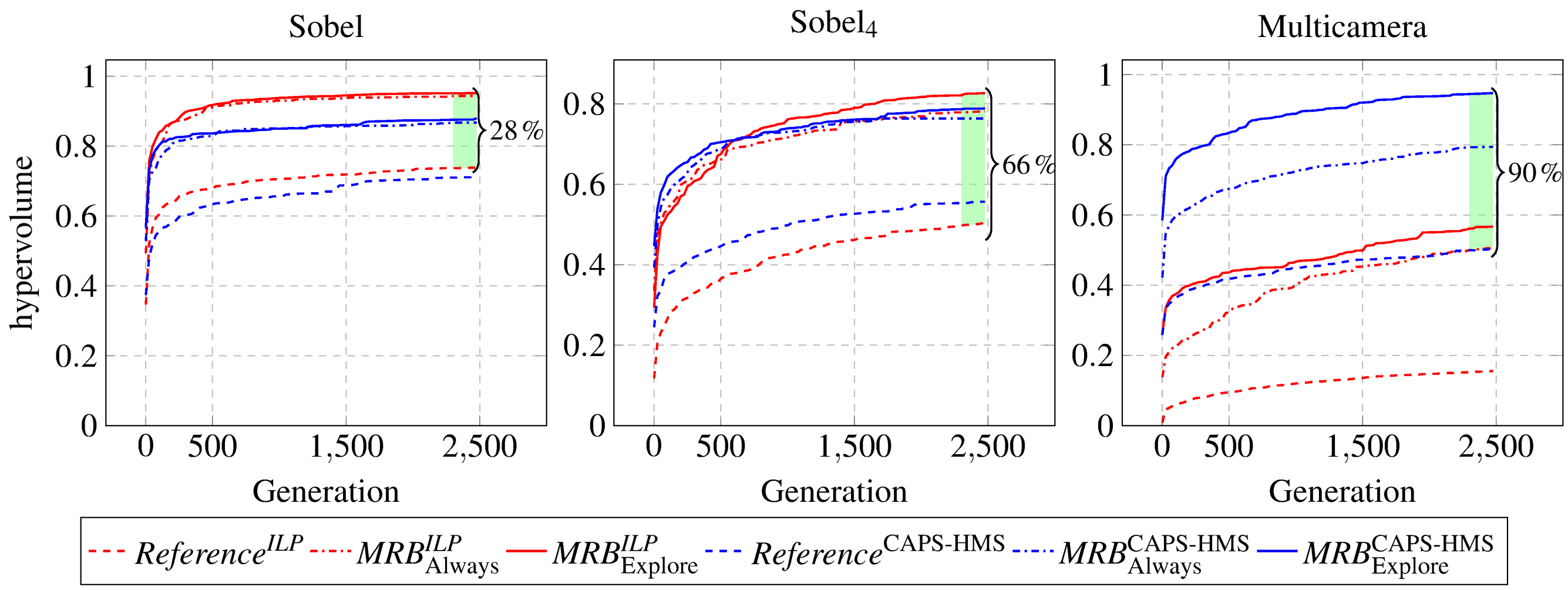

**FIGURE 8.** Hypervolume scores obtained for the presented applications over 2,500 generations for the six approaches.

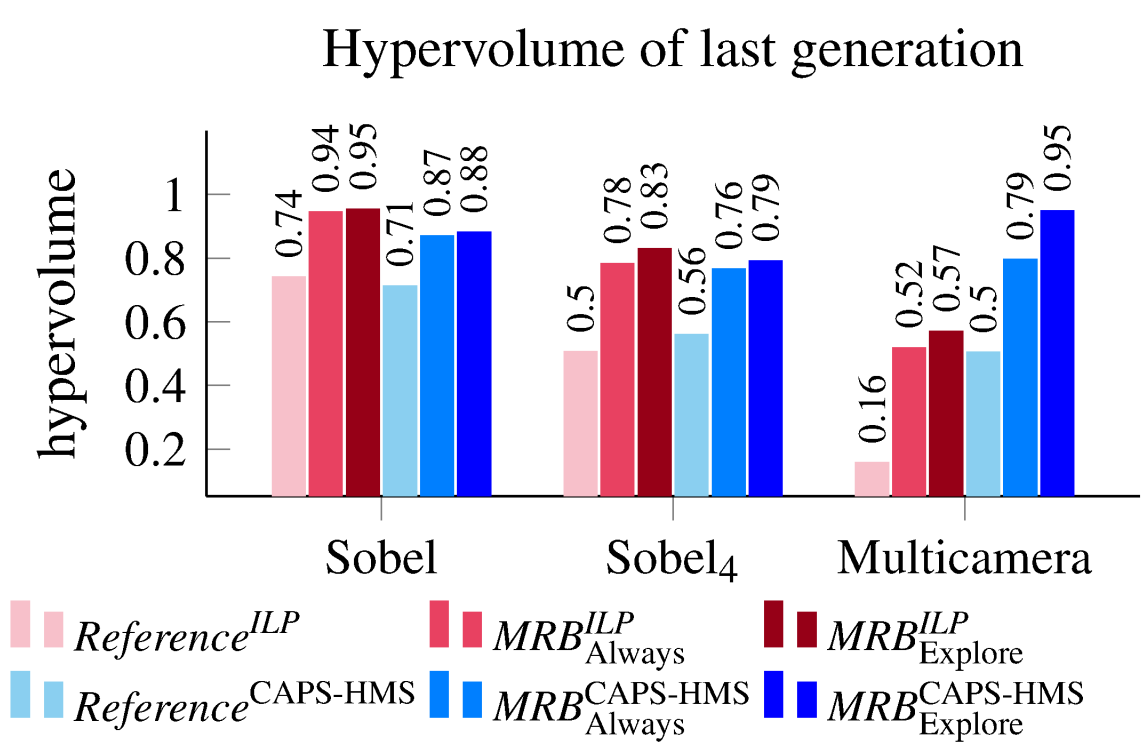

**FIGURE 9.** Hypervolume obtained for the presented applications of the last explored generation for the six approaches.

To alleviate this, we might increase the timeout of the ILP, but this would be detrimental to the exploration runtime. Detailed explanations follow from an exploration time analysis in Section VI-B. We conclude that only if the specification is relatively small, the ILP-based $MRB^{ILP}_{\text{Explore}}$ approach may find slightly better Pareto-front approximations after an equal number of generations, but it provides inferior solutions for larger problem sizes where the efficient $MRB^{\text{CAPS-HMS}}_{\text{Explore}}$ approach is by far superior.

Next, we examine the Pareto fronts for each approach and application to more closely check whether replacing multi-cast actors with MRBs will always provide superior solutions. It will also be shown how improvements in the hypervolume score are reflected in the space of the design objectives. For each of the three applications, the union of Pareto fronts achieved in the last generation by the strategies *Reference*, $MRB_{\text{Always}}$, and $MRB_{\text{Explore}}$ using the ILP-based, respectively, the CAPS-HMS decoder are shown in Figs. 10 and 11. Each plot in these figures displays the objective space spanned by period $\mathbf{P}$, memory footprint $\mathbf{M_F}$, and core cost $\mathbf{K}$. The core cost $\mathbf{K}$ is thereby represented by a color gradient from blue to red (as shown in the color map on the right). Circle, square, and triangle symbols represent solutions obtained by the strategies *Reference*, $MRB_{\text{Always}}$, and $MRB_{\text{Explore}}$. Filled symbols indicate non-dominated solutions from the union of the Pareto fronts of all three approaches combined.

**Key Observations:** First, most filled symbols present in each plot correspond to the strategy $MRB_{\text{Explore}}$, regardless of whether the ILP or the CAPS-HMS decoder is used, leading to the conclusion that most of the non-dominated solutions are found when using the strategy $MRB_{\text{Explore}}$, which selectively replaces multi-cast actors with MRBs.

Second, for each application, when comparing circle to triangle symbols denoting comparable period solutions, circle symbols indicate a much larger memory footprint $\mathbf{M_F}$ than the corresponding triangle symbols. Hence, the selective replacement of multi-cast actors by MRBs, in contrast to not including any MRB, leads to significant savings in memory footprint. For example, most of the non-dominated points for the multicamera application shown in Fig. 11 for the approach $MRB^{\text{CAPS-HMS}}_{\text{Explore}}$ require between 32 and 40 MiB of memory (see filled triangles). In contrast, the memory footprint of the non-dominated solutions found by the $Reference^{\text{CAPS-HMS}}$ approach (see circle symbols) vary between 55 and 90 MiB.

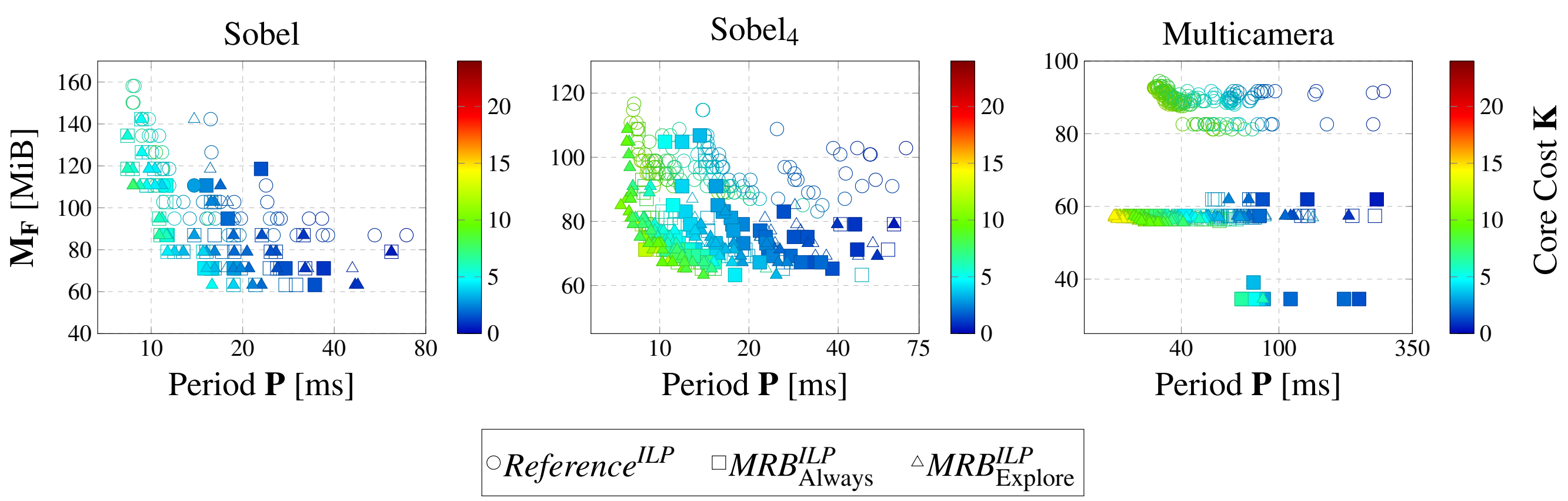

**FIGURE 10.** Union of the Pareto fronts of the last generation obtained for the presented applications after 2,500 generations using the **ILP-based** decoder. Filled points are non-dominated solutions of the union of the three Pareto fronts. The period **P** is presented in a logarithmic scale for better visualization.

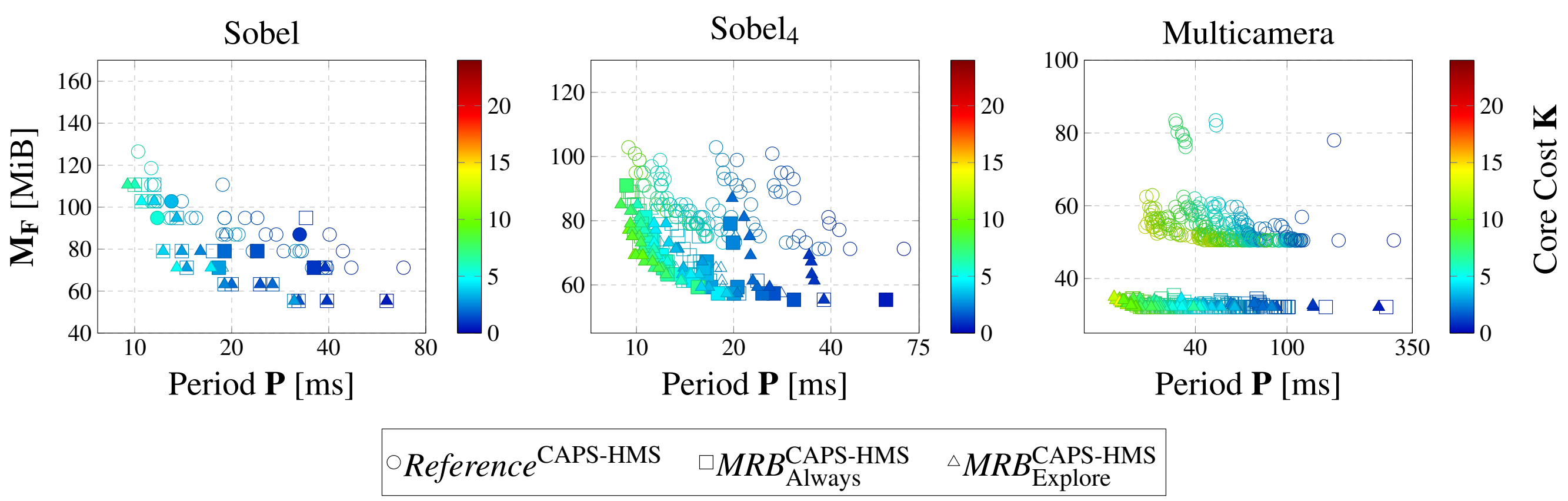

**FIGURE 11.** Union of the Pareto fronts of the last generation obtained for the presented applications after 2,500 generations using the **heuristic-based** (CAPS-HMS) decoder. Filled points are non-dominated solutions of the union of the three Pareto fronts. The period **P** is presented in a logarithmic scale for better visualization.

Third, the shortest-period solution for the $\text{Sobel}_4$ and multicamera applications are characterized by a filled triangle symbol. Moreover, when examining shortest-period solutions for a given memory footprint, one can observe that almost all of these are found by the strategy $MRB_{\text{Explore}}$ (filled triangles). This validates our assertion from Section III-D that there are cases where an MRB replacement is detrimental to (i.e., it increases) the execution period. Thus, we can conclude that for mid to large size applications containing a non-negligible number of multi-cast actors, the selective replacement of these multi-cast actors by MRBs may lead to shorter periods compared to not including any MRB or replacing all multi-cast actors with MRBs.

## B. EXPLORATION TIME

Another essential feature for evaluating a DSE approach is the exploration time. In the context of DSE, the evaluation time is crucial because a DSE run may require thousands of design point evaluations [21]. Table 2 presents the exploration time in seconds when using the ILP decoder and the CAPS-HMS decoder after 2,500 generations. We also present the speedup ratio, comparing the time of the much faster heuristic-based CAPS-HMS decoder against the ILP. The speedup for each DSE approach is calculated as follows:

$$\begin{aligned} &\mathsf{DSE} \in \{Reference, MRB_{\text{Always}}, MRB_{\text{Explore}}\} \\ &\text{speedup}(\mathsf{DSE}) = \frac{\text{runtime}(\mathsf{DSE}^{ILP})}{\text{runtime}(\mathsf{DSE}^{\text{CAPS-HMS}})} \end{aligned} \tag{28}$$

**Key Observations:** In general, we can observe that the ILP-based decoder requires significantly more time to perform the exploration of the design space for a given number of generations when compared to the CAPS-HMS decoder. Even for the small Sobel application, the ILP-based approach $Reference^{ILP}$ takes 16.02 hours to complete 2,500 generations, while the approach $Reference^{\text{CAPS-HMS}}$ only requires 7.38 minutes. The reported speedup range of CAPS-HMS is 125× to 149× for the Sobel application. However, both approaches require more time to explore middle- to large-size applications. For $\text{Sobel}_4$, the CAPS-HMS decoder requires between 27.19 and 42.28 minutes, while the ILP decoder requires between 20 and 22.68 hours to perform 2,500 generations of the optimization loop. Accordingly, for the $Sobel_4$, the speedup range of the CAPS-HMS decoder is between 28× and 50×. For the largest multicamera application, the exploration time varies between 1.60 and 4.54 hours for CAPS-HMS. In contrast, the ILP decoder takes between 14.62 and 17.80 hours. There, the reported speedup of CAPS-HMS ranges between 4× and 9×. Note here that the reported speedup range is lower compared to the other applications because the ILP-based decoder has a timeout of three seconds. In summary, the ILP-based decoder is best suited for small to mid-size applications, as it is then able to find a minimal period schedule for any given binding of actors to cores and channels to memories. In contrast, the proposed CAPS-HMS heuristic is the preferable solution for realistically sized applications, as solving explodes with an increasing number of variables.

**TABLE 2.** Exploration time[6] comparison of CAPS-HMS decoder against the ILP decoder for running 2,500 generations.

| Application | runtime ILP [hours] | | | runtime CAPS-HMS [minutes] | | | speedup | | |
|---|---|---|---|---|---|---|---|---|---|
| | $Reference^{ILP}$ | $MRB^{ILP}_{Always}$ | $MRB^{ILP}_{Explore}$ | $Reference^{\text{CAPS-HMS}}$ | $MRB^{\text{CAPS-HMS}}_{Always}$ | $MRB^{\text{CAPS-HMS}}_{Explore}$ | $Reference$ | $MRB_{Always}$ | $MRB_{Explore}$ |
| Sobel | 16.02 | 12.00 | 13.30 | 7.38 | 5.76 | 5.37 | 130× | 125× | 149× |
| $Sobel_4$ | 20.00 | 22.68 | 20.84 | 42.28 | 27.19 | 31.53 | 28× | 50× | 40× |
| Multicamera | 17.80 | 15.61 | 14.62 | 272.87 | 137.77 | 96.32 | 4× | 7× | 9× |

## VII. RELATED WORK

Approaches for optimizing parallel implementation of applications specified as dataflow networks [22] perform multi-objective optimization of conflicting design objectives, e.g., throughput and number of allocated cores. On the one hand, approaches such as [15, 23] optimize dataflow applications' throughput and the number of allocated cores in a given architecture. However, the previously presented approaches do not consider any memory footprint evaluation of implementations or the generation of periodic schedules during DSE.

In the following, we categorize the related work as approaches performing memory footprint minimization and approaches generating periodic schedules.

### A. MEMORY FOOTPRINT MINIMIZATION

Approaches for memory footprint minimization can be classified into two main categories: (i) approaches minimizing the size of FIFOs and (ii) approaches implementing memory-reuse strategies that allow different FIFOs to be mapped into overlapping memory spaces or track individual token lifetimes to exploit memory footprint reductions over the execution of an application. In the first category, techniques such as FIFO sizing have been widely studied to reduce the memory footprint of Synchronous Dataflow (SDF) applications [24–26]. Such approaches determine the minimal buffer size of an SDF application under throughput constraints. However, those approaches do not consider any memory-reuse strategy because each buffer is studied as a separate unit allocated in memory, and no shared memory address space is considered. In the second category, the approach presented in [27] derives overlapping memory allocations for individual tokens communicated during the execution of an SDF graph. However, it assumes no overlap between iterations, i.e., an execution period only contains actor firings of a single iteration. Thus, the achievable minimal period is severely constrained. Apart from performing an agnostic memory footprint minimization, some approaches exploit the knowledge about the application and actor characteristics. For instance, dataflow frameworks [8, 12, 28] targeting image processing apply memory minimization strategies based on the behavior of a set of specialized actors performing operations like multi-cast, fork, and join of data. For instance, the employed memory minimization strategy described in [12] merges all outgoing buffers of a multi-cast actor by replacing them with a broadcast FIFO that supports a single writer but multiple readers [12]. However, no other design objectives apart from memory footprint are explored. In this paper, we propose a holistic approach that considers not only the minimization of memory footprint but also the mapping and scheduling of communication channels and actors onto heterogeneous many-core architectures as well as the number of allocated CPUs as exploration objectives.

### B. SCHEDULING

There exist approaches for communication-aware scheduling of Directed Acyclic Graphs (DAGs) targeting many-cores that can be classified according to the utilized scheduling method: heuristic-based – i.e., list-scheduling [29] and clustering-scheduling [30] – or meta-heuristics-based – i.e., genetic algorithms [31–33], simulated annealing [34], and particle swarm [35] –, to mention a few. Although able to take into account communication scheduling, the optimization goal is to minimize the schedule make-span, i.e., the latency of a single iteration. Thus, minimum periodic schedules are not achievable by the mentioned approaches. Moreover, the communications on the DAGs are often not explicitly specified, but rather using a Communication-to-Computation Ratio (CCR), i.e., no explicit communications over interconnect resources in the target architecture are modeled.

When analyzing dataflow, scheduling strategies applied at compile time are beneficial [15]. E.g., Self-Timed Execution (STE) [24, 36] simulates the execution of a dataflow graph by using so-called state transformations. The state of a DFG is encoded as a set of variables representing the current state of the system. Changes during the execution of a system – e.g., an actor consuming/producing tokens from/to a channel – are represented by state transformations. During the simulation of the system, the transforming states are recorded until a periodic pattern emerges, which corresponds to the periodic schedule of the DFG. However, STE does not consider any communication in the scheduling. As a remedy, [37, 38] proposed an extension to STE by including communication delay in the model. However, these works can only achieve schedules targeting MPSoC architectures with a single bus and a global memory. Thus, the model assumes a single resource to schedule the communication at a fixed bandwidth. This is different to our approach, which is able to target heterogeneous many-core architectures composed of a hierarchical organization of cores, memories, and interconnects.

[6] The exploration times shown in Table 2 were obtained by running the presented DSE approaches on a 4-core Intel(R) Core(TM) i7-4790 running at 3.60 GHz with 32 GB of RAM.

Last, *modulo-scheduling* is a well-known loop scheduling technique applied in compiler optimizations as well as to periodic scheduling of DAGs on fine [39–43] and coarse-grained architectures [44–47]. There, applications are modeled as DAGs, and hardware units such as adders, multipliers, and accelerators are used to modulo-schedule a hardware implementation of an iterative application [43]. E.g., approaches such as [39, 42] used modulo scheduling in combination with loop unrolling during high-level synthesis. Approaches such as [45, 47] perform loop unrolling of applications composed of tasks mapped to the processing elements of coarse-grained architectures. However, these approaches ignore the scheduling of communications, i.e., transfers of data from cores to memory and from memories to cores over communication resources such as buses or NoCs.

This paper considered an explorative approach to map and schedule dataflow specifications on heterogeneous multi-core architectures by considering as well the scheduling of actors as the communications between actors. Our approach targets heterogeneous many-core architectures where cores of different kinds might exist in the same architecture, and complex communications are explicitly modeled, mapped, and scheduled on interconnect resources and memories respecting a hierarchical tile organization. As illustrated, the mapping of as well actors to cores as data buffers in channels to memories, including processor-local memory, tile-local memory, and global memory, is explored during a DSE. For each solution candidate, a periodic schedule is then optimized either using an ILP formulation or an efficient scheduling heuristic called CAPS-HMS.

## VIII. CONCLUSIONS

As a first contribution, this paper introduces the concept of Multi-Reader Buffers (MRBs) as a memory-efficient implementation of multi-cast actors and their replacement as a graph transformation. Rather than replicating produced tokens for all readers, an MRB stores only one token, which is alive until the last reader has consumed it. MRB replacement provides minimal buffer implementations obtained by replacing all multi-cast actors in an application with MRBs. However, replacing multi-cast actors with MRBs may increase the execution period – i.e., reduce the throughput – due to communication contention when accessing shared data.

To properly examine these trade-offs, as our second contribution, we propose a multi-objective Design Space Exploration (DSE) approach that selectively decides the replacement of multi-cast actors with MRBs and explores FIFO and channel mappings to trade memory footprint, core cost, and period of schedules. It is shown that the quality of found solutions improves when selectively replacing multi-cast actors with MRBs within a range of 28 % to 90 % in solution quality measured by a hypervolume indicator.

Moreover, as our third contribution, we proposed and compared two scheduling approaches that are used to determine a periodic schedule for the actors as well as the read/write accesses to buffers for each explored design point during the DSE: First, an ILP formulation that delivers the exact minimum period given an application binding. This ILP formulation performs well in terms of solution times for small to mid-sized applications. The second is a fast CAPS-HMS heuristic approach that performs particularly well when tackling large applications. It has been shown that for the small and mid-sized applications used in the experiments, our proposed CAPS-HMS is only slightly inferior by 7 % in terms of hypervolume compared to the ILP. But for large applications and the complexity of the target architecture, the ILP solution times can become prohibitively long. In contrast, the fast CAPS-HMS outperforms the ILP by 67 % in hypervolume for our largest test application. Finally, the presented DSE approach is distinguished from the state-of-the-art by considering (i) constraints in the memory size of each on-chip-memory, (ii) memory hierarchies, (iii) support of heterogeneous many-core platforms, and (iv) optimization of buffer placement and overall scheduling to minimize the period.

## REFERENCES

[1] E. Lee, "The Problem with Threads," *Computer*, vol. 39, no. 5, pp. 33–42, 2006.

[2] J. Dennis, "First Version of a Data Flow Procedure Language," in *Programming Symposium*, ser. Lecture Notes in Computer Science, B. Robinet, Ed. Berlin, Heidelberg: Springer-Verlag, 1974, vol. 19, pp. 362–376.

[3] F. Commoner, A. W. Holt, S. Even, and A. Pnueli, "Marked Directed Graphs," *Journal of Computer and System Sciences*, vol. 5, no. 5, pp. 511–523, Oct. 1971.

[4] J. Falk, K. Neubauer, C. Haubelt, C. Zebelein, and J. Teich, *Integrated Modeling Using Finite State Machines and Dataflow Graphs*. Cham: Springer International Publishing, 2019, pp. 825–864.

[5] T. Blickle, J. Teich, and L. Thiele, "System-level synthesis using evolutionary algorithms," *Design Automation for Embedded Systems*, vol. 3, no. 1, pp. 23–58, 1998.

[6] M. Letras, J. Falk, S. Wildermann, and J. Teich, "Automatic Conversion of Simulink Models to SysteMoC Actor Networks," in *Proc. of SCOPES*. New York, NY, USA: ACM, 2017, pp. 81–84.

[7] J. Keinert, M. Streubühr, T. Schlichter, J. Falk, J. Gladigau, C. Haubelt, J. Teich, and M. Meredith, "SYSTEMCODESIGNER - an Automatic ESL Synthesis Approach by Design Space Exploration and Behavioral Synthesis for Streaming Applications," *ACM Trans. on Design Automation of Electronic Systems*, vol. 14, no. 1, pp. 1:1–1:23, Jan. 2009.

[8] K. Desnos, M. Pelcat, J.-F. Nezan, and S. Aridhi, "On Memory Reuse Between Inputs and Outputs of Dataflow Actors," *ACM Transactions on Embedded Computing Systems*, vol. 15, no. 2, Feb. 2016.

[9] M. Letras, J. Falk, and J. Teich, "Throughput and Memory Optimization for Parallel Implementations of Dataflow Networks Using Multi-Reader Buffers," in *Fourth Workshop on Next Generation Real-Time Embedded Systems (NG-RES 2023)*, ser. Open Access Series in Informatics (OASIcs), F. Terraneo and D. Cattaneo, Eds., vol. 108. Dagstuhl, Germany: Schloss Dagstuhl – Leibniz-Zentrum für Informatik, 2023, pp. 6:1–6:13.

[10] J. Teich, "Hardware/software codesign: The past, the present, and predicting the future," *Procs. of the IEEE*, vol. 100, no. Special Centennial Issue, pp. 1411–1430, May 2012.

[11] S. Ha and J. Teich, *Handbook of Hardware/Software Codesign*, 1st ed. Springer Publishing Company, Incorporated, 2017.

[12] A. R. Mamidala, D. Faraj, S. Kumar, D. Miller, M. Blocksome, T. Gooding, P. Heidelberger, and G. Dozsa, "Optimizing mpi collectives using efficient intra-node communication techniques over the blue gene/p supercomputer," in *2011 IEEE International Symposium on Parallel and Distributed Processing Workshops and Phd Forum*, 2011, pp. 771–780.

[13] OpenDSE, "Open Design Space Exploration Framework," *http://opendse.sf.net/*, 2018.

[14] M. Lukasiewycz, M. Glaß F. Reimann, and J. Teich, "Opt4J: A Modular Framework for Meta-heuristic Optimization," in *GECCO*, ser. GECCO '11. New York, NY, USA: ACM, 2011, pp. 1723–1730.

[15] M. Letras, J. Falk, T. Schwarzer, and J. Teich, "Multi-objective Optimization of Mapping Dataflow Applications to MPSoCs Using a Hybrid Evaluation Combining Analytic Models and Measurements," *ACM Trans. on Design Automation of Electronic Systems*, vol. 26, p. 1–33, 2020.

[16] T. H. Cormen, C. E. Leiserson, R. L. Rivest, and C. Stein, *Introduction to algorithms*. MIT press, 2022.

[17] K. Deb, A. Pratap, S. Agarwal, and T. Meyarivan, "A Fast and Elitist Multiobjective Genetic Algorithm: NSGA-II," *Trans. Evol. Comp*, vol. 6, no. 2, pp. 182–197, Apr. 2002.

[18] A. P. Guerreiro, C. M. Fonseca, and L. Paquete, "The hypervolume indicator: Computational problems and algorithms," *ACM Comput. Surv.*, vol. 54, no. 6, jul 2021.

[19] E. Zitzler and L. Thiele, "Multiobjective optimization using evolutionary algorithms — a comparative case study," in *Parallel Problem Solving from Nature — PPSN V*, A. E. Eiben, T. Bäck, M. Schoenauer, and H.-P. Schwefel, Eds. Berlin, Heidelberg: Springer Berlin Heidelberg, 1998, pp. 292–301.

[20] K. Ciesielski, "How good is lebesgue measure?" *The Mathematical Intelligencer*, vol. 11, pp. 54–58, 1989.

[21] A. D. Pimentel, "Exploring exploration: A tutorial introduction to embedded systems design space exploration," *IEEE Design Test*, vol. 34, no. 1, pp. 77–90, Feb 2017.

[22] L. Thiele, K. Strehl, D. Ziegengein, R. Ernst, and J. Teich, "Funstate-an internal design representation for codesign," in *1999 IEEE/ACM International Conference on Computer-Aided Design. Digest of Technical Papers (Cat. No.99CH37051)*, 1999, pp. 558–565.

[23] J. Falk, J. Keinert, C. Haubelt, J. Teich, and S. Bhattacharyya, "A Generalized Static Data Flow Clustering Algorithm for MPSoC Scheduling of Multimedia Applications," in *Proceedings of ACM International Conference on Embedded Software*, Oct. 2008, pp. 189–198.

[24] S. Stuijk, M. Geilen, and T. Basten, "Exploring trade-offs in buffer requirements and throughput constraints for synchronous dataflow graphs," in *Proceedings of Design Automation Conference (DAC)*, 2006, pp. 899–904.

[25] M. Benazouz, O. Marchetti, A. Munier-Kordon, and P. Urard, "A new approach for minimizing buffer capacities with throughput constraint for embedded system design," in *ACS/IEEE International Conference on Computer Systems and Applications (AICCSA)*, 2010, pp. 1–8.

[26] Q. Tang, T. Basten, M. Geilen, S. Stuijk, and J.-B. Wei, "Task-fifo co-scheduling of streaming applications on mpsocs with predictable memory hierarchy," *ACM Trans. Embed. Comput. Syst.*, vol. 16, no. 2, mar 2017.

[27] K. Desnos, M. Pelcat, J.-F. Nezan, and S. Aridhi, "Memory analysis and optimized allocation of dataflow applications on shared-memory mpsocs," *Journal of Signal Processing Systems*, vol. 80, no. 1, pp. 19–37, 2015.

[28] H. Yviquel, A. Sanchez, P. Jääskeläinen, J. Takala, M. Raulet, and E. Casseau, "Embedded multi-core systems dedicated to dynamic dataflow programs," *Journal of Signal Processing Systems*, vol. 80, no. 1, pp. 121–136, 2015.

[29] H. Wang and O. Sinnen, "List-scheduling versus cluster-scheduling," *IEEE Transactions on Parallel and Distributed Systems*, vol. 29, no. 8, pp. 1736–1749, 2018.

[30] N. Shah, W. Meert, and M. Verhelst, "Graphopt: Constrained-optimization-based parallelization of irregular graphs," *IEEE Transactions on Parallel and Distributed Systems*, vol. 33, no. 12, 2022.

[31] M. Akbari, H. Rashidi, and S. H. Alizadeh, "An enhanced genetic algorithm with new operators for task scheduling in heterogeneous computing systems," *Engineering Applications of Artificial Intelligence*, vol. 61, pp. 35–46, 2017.

[32] Y. Xu, K. Li, J. Hu, and K. Li, "A genetic algorithm for task scheduling on heterogeneous computing systems using multiple priority queues," *Information Sciences*, vol. 270, pp. 255–287, 2014.

[33] E. C. da Silva and P. H. R. Gabriel, "A comprehensive review of evolutionary algorithms for multiprocessor dag scheduling," *Computation*, vol. 8, no. 2, 2020.

[34] A. Saad, A. Kafafy, O. A. El Raouf, and N. El-Hefnawy, "A grasp-simulated annealing approach applied to solve multi-processor task scheduling problems," in *2019 14th International Conference on Computer Engineering and Systems (ICCES)*, 2019, pp. 310–315.

[35] M. Gao, Y. Zhu, and J. Sun, "The multi-objective cloud tasks scheduling based on hybrid particle swarm optimization," in *2020 Eighth International Conference on Advanced Cloud and Big Data (CBD)*, 2020, pp. 1–5.

[36] X.-Y. Zhu, M. Geilen, T. Basten, and S. Stuijk, "Multiconstraint static scheduling of synchronous dataflow graphs via retiming and unfolding," *IEEE Transactions on Computer-Aided Design of Integrated Circuits and Systems*, vol. 35, no. 6, pp. 905–918, 2016.

[37] M. Ma and R. Sakellariou, "Communication-aware scheduling algorithms for synchronous dataflow graphs on multicore systems," in *Proceedings of the 18th International Conference on Embedded Computer Systems: Architectures, Modeling, and Simulation*, ser. SAMOS '18. New York, NY, USA: Association for Computing Machinery, 2018, p. 55–64.

[38] ——, "Code-size-aware scheduling of synchronous dataflow graphs on multicore systems," *ACM Trans. Embed. Comput. Syst.*, vol. 20, no. 3, mar 2021.

[39] J. Oppermann, A. Koch, M. Reuter-Oppermann, and O. Sinnen, "Ilp-based modulo scheduling for high-level synthesis," in *Proceedings of the International Conference on Compilers, Architectures and Synthesis for Embedded Systems*, ser. CASES '16. New York, NY, USA: Association for Computing Machinery, 2016.

[40] P. Sittel, N. Fiege, J. Wickerson, and P. Zipf, "Optimal and heuristic approaches to modulo scheduling with rational initiation intervals in hardware synthesis," *IEEE Transactions on Computer-Aided Design of Integrated Circuits and Systems*, vol. 41, no. 3, pp. 614–627, 2022.

[41] N. Fiege and P. Zipf, "Bloop: Boolean satisfiability-based optimized loop pipelining," *ACM Trans. Reconfigurable Technol. Syst.*, vol. 16, no. 3, jul 2023.

[42] J. Oppermann, M. Reuter-Oppermann, L. Sommer, A. Koch, and O. Sinnen, "Exact and practical modulo scheduling for high-level synthesis," *ACM Trans. Reconfigurable Technol. Syst.*, vol. 12, no. 2, may 2019.

[43] J. Teich, "Synthesis and optimization of digital hardware/software systems," in *System Design Automation: Fundamentals, Principles, Methods, Examples*. Springer, 1996, pp. 3–26.

[44] R. Ferreira, V. Duarte, W. Meireles, M. Pereira, L. Carro, and S. Wong, "A just-in-time modulo scheduling for virtual coarse-grained reconfigurable architectures," in *2013 International Conference on Embedded Computer Systems: Architectures, Modeling, and Simulation (SAMOS)*, 2013, pp. 188–195.

[45] M. Witterauf, A. Tanase, F. Hannig, and J. Teich, "Modulo scheduling of symbolically tiled loops for tightly coupled processor arrays," in *2016 IEEE 27th International Conference on Application-specific Systems, Architectures and Processors (ASAP)*, 2016, pp. 58–66.

[46] H. Park, K. Fan, S. A. Mahlke, T. Oh, H. Kim, and H.-s. Kim, "Edge-centric modulo scheduling for coarse-grained reconfigurable architectures," in *Proceedings of the 17th International Conference on Parallel Architectures and Compilation Techniques*, ser. PACT '08. New York, NY, USA: Association for Computing Machinery, 2008, p. 166–176.

[47] C. Tirelli, L. Ferretti, and L. Pozzi, "Sat-mapit: A sat-based modulo scheduling mapper for coarse grain reconfigurable architectures," in *2023 Design, Automation & Test in Europe Conference & Exhibition (DATE)*. IEEE, 2023, pp. 1–6.